\documentclass{sigchi}

\toappear{\scriptsize Permission to make digital or hard copies of all or part of this work for personal or classroom use is granted without fee provided that copies are not made or distributed for profit or commercial advantage and that copies bear this notice and the full citation on the first page. Copyrights for components of this work owned by others than the author(s) must be honored. Abstracting with credit is permitted. To copy otherwise, or republish, to post on servers or to redistribute to lists, requires prior specific permission and/or a fee. Request permissions from permissions@acm.org. \\
{\emph{UIST '20, October 20--23, 2020, Virtual Event, USA} } \\
\copyright~2020 Association for Computing Machinery. \\
ACM ISBN 978-1-4503-7514-6/20/10\ ...\$15.00. \\
http://dx.doi.org/10.1145/3379337.3415858}

\usepackage{balance}       
\usepackage{graphics}      
\usepackage[T1]{fontenc}   
\usepackage{txfonts}
\usepackage{mathptmx}
\usepackage[pdflang={en-US},pdftex]{hyperref}
\usepackage{color}
\usepackage{booktabs}
\usepackage{textcomp}
\usepackage{enumitem}

\usepackage{multirow,graphicx}

\usepackage{etoolbox}
\makeatletter
\patchcmd{\@verbatim}
  {\verbatim@font}
  {\verbatim@font\small}
  {}{}
\makeatother

\usepackage{microtype}        
\usepackage{ccicons}          

\usepackage{todonotes}
\usepackage{csquotes}
\usepackage{epigraph}

\def\plaintitle{PolicyKit: Building Governance in Online Communities}

\def\emptyauthor{}
\def\plainkeywords{governance; policy; toolkit; moderation; online communities}

\makeatletter
\def\url@leostyle{%
  \@ifundefined{selectfont}{
    \def\UrlFont{\sf}
  }{
    \def\UrlFont{\small\bf\ttfamily}
  }}
\makeatother
\def\pprw{8.5in}
\def\pprh{11in}

\definecolor{linkColor}{RGB}{6,125,233}
\hypersetup{%
  pdftitle={\plaintitle},
  pdfauthor={\emptyauthor},
  pdfkeywords={\plainkeywords},
  pdfdisplaydoctitle=true, 
  bookmarksnumbered,
  pdfstartview={FitH},
  colorlinks,
  citecolor=black,
  filecolor=black,
  linkcolor=black,
  urlcolor=linkColor,
  breaklinks=true,
  hypertexnames=false
}

\newcommand\Mark[1]{\textsuperscript#1}

\begin{document}

\title{\plaintitle}

\numberofauthors{3}

    \author{
           Amy X. Zhang\textsuperscript{1,2},  
           Grant Hugh\Mark{2},    
           Michael S. Bernstein\Mark{2}
          \end{tabular}
        \centering
        \begin{tabular}{ p{20em}  p{20em} }
        \hfil\affaddr{University of Washington\Mark{1}} &
          \hfil\affaddr{Stanford University\Mark{2}} \\
          \hfil\affaddr{Seattle, WA} & \hfil\affaddr{Stanford, CA}\\
          \hfil\email{axz@cs.uw.edu} & \hfil\email{ghugh@stanford.edu,}
          \email{msb@cs.stanford.edu}\\
    }

\maketitle

\begin{abstract}

The software behind online community platforms encodes a governance model that represents a strikingly narrow set of governance possibilities focused on moderators and administrators. 
When online communities desire other forms of government, such as ones that take many members' opinions into account or that distribute power in non-trivial ways, communities must resort to laborious manual effort. 
In this paper, we present \mbox{PolicyKit}, a software infrastructure that empowers online community members to concisely author a wide range of governance procedures and automatically carry out those procedures on their home platforms. 
We draw on political science theory to encode community governance into policies, or short imperative functions that specify a procedure for determining whether a user-initiated action can execute. 
Actions that can be governed by policies encompass everyday activities such as posting or moderating a message, but actions can also encompass changes to the policies themselves, enabling the evolution of governance over time. We demonstrate the expressivity of PolicyKit through implementations of governance models such as a random jury deliberation, a multi-stage caucus, a reputation system, and a promotion procedure inspired by Wikipedia's Request for Adminship (RfA) process.

\end{abstract}

\ccsdesc[500]{Human-centered computing~Collaborative and social computing systems and tools}

\printccsdesc

\keywords{\plainkeywords}

\section{Introduction}

\begin{displayquote}
\textit{``In democratic countries the science of association is the mother science; the progress of all the others depends on the progress of that one.''}--Alexis de Tocqueville, 1835~\cite{de2003democracy}
\end{displayquote}

Millions of communities gather in online spaces such as Slack workspaces, Reddit subreddits, Facebook groups, and mailing lists. 
These communities fill an important part of our everyday lives~\cite{kraut2012building} and go on to shape broader social institutions~\cite{noveck2009wiki}. 
As a result, decisions around the \textit{governance} of online communities, such as who can join, what content is allowed, and the consequences for breaking rules, have great importance. 
Today, this governance is predominantly expressed as a model consisting of roles and permissions, where groups such as administrators and moderators have broad privileges over regular users. 
This roles-and-permissions model has its roots in the UNIX file permissions model developed nearly fifty years ago~\cite{schneider-admins2019}, and it is a model now enshrined within the \textit{software} of almost all major community platforms.
 
But governance via roles and permissions describes only a narrow set of governance possibilities. This approach also encodes certain values, making it easier to implement governance that is top-down, autocratic, and punitive~\cite{frey2019emergence,schoenebeck2020drawing,Shaw2014Oligarchy}. 
There are many cases where an online community may prefer or be better served by a different style of governance. 
For instance, the English Wikipedia community has chosen to make major decisions through a deliberative process where all users can provide input~\cite{burke2008mopping,im2018deliberation}; the node.js project follows a consensus-seeking decision model~\cite{nodejs}; and Slashdot practices meta-moderation where the moderators themselves get reviewed~\cite{lampe2004slash}. 
Each of these processes embodies the particular flavor of the participatory values of its community~\cite{kelty2018two}.

Unfortunately, the software underlying most community platforms cannot support these alternative forms of government. Wikipedians must carry out their deliberative procedure manually, with some help from custom-written bots, instead of relying on MediaWiki software, which itself only embeds a permissions model.
Manually carrying out governance procedures is cumbersome and error-prone. In addition, 
community members may not be aware of policies or choose not to comply with them~\cite{butler2008don,matias2019preventing}.
Given the difficulty involved, it is no surprise that most online communities use their platform's default permissions model, even when it may not be a good fit. Without the flexibility to articulate new governance models, communities have few options when contending with problems common to moderated communities, ranging from moderator burnout~\cite{dosono2019moderation,gillespie2018custodians,roberts2019behind}, to being overwhelmed by newcomers~\cite{kiene2016surviving}, to surviving a legitimacy crisis~\cite{cohen2007contributor}.

\begin{figure*}
  \centering
  \includegraphics[width=2\columnwidth]{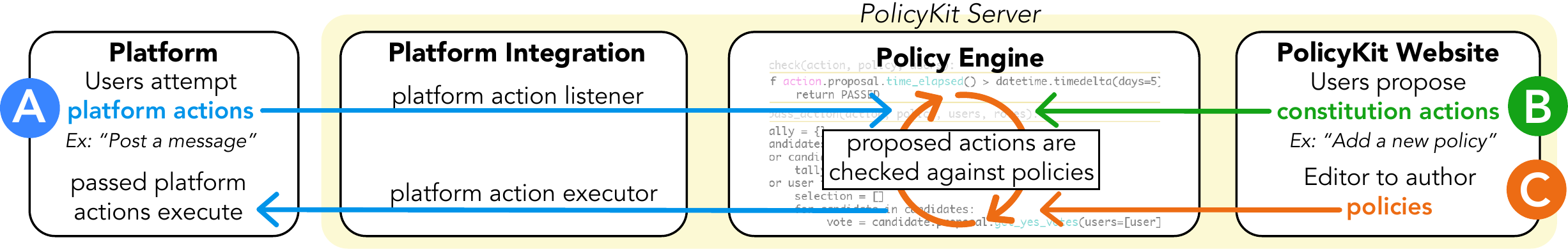}
  \caption{Rather than following a permissions model, PolicyKit represents governance as a set of \textit{procedures} that are used to make decisions. (A) Users attempt everyday actions, such as posting a message, on their community platform. These actions are caught by PolicyKit's action listener and checked against policies in the policy engine. Each policy articulates an action scope and a procedure for determining whether those action can pass, for example, a vote by platform members. If the action passes, it executes back on the platform. 
  (B) Users initiate constitution
   actions that alter the community's governance policies. 
   These also go to the policy engine and execute there if passed. 
  (C) Users can author policies within an editor in the PolicyKit website. 
  If the policy passes (via a constitution action), it becomes one of the policies that the policy engine considers when checking actions.}~\label{fig:concept}
\end{figure*}

In this work, we present \textit{PolicyKit}, a software infrastructure that empowers online communities to create a broad range of governance approaches by writing a small amount of code. 
This governance is then automatically carried out on behalf of the community on the community's home platform such as Reddit or Slack.
Unlike a standard permissions model, where an action such as ``\textit{post a message to the group}'' depends on the permissions granted to the user performing the action, PolicyKit enables community members to author short executable scripts, or \textit{policies}, that specify a procedure---a set of steps to follow---for determining whether an action can execute. 
A policy might, for example, require that three random community members review the message and approve it through a majority vote.
Drawing from Ostrom's Institutional Analysis and Development (IAD)~\cite{ostrom}, a theoretical framework for describing governance arrangements, PolicyKit's main insight is to shift governance from articulating \textit{permissions} to articulating \textit{procedures}, where procedures can express a wide range of governance models concisely, including participatory and democratic models. 
And since policies are written in code, PolicyKit can execute them as specified without requiring manual effort from users each time.
The actions that can be governed by policies include everyday actions that take place on a platform, such as posting a message, as well as actions that alter the governance model itself, such as introducing a new policy (Figure~\ref{fig:concept}). This allows communities to evolve their governance over time using PolicyKit.

We envision that PolicyKit will allow communities to iteratively develop governance of their own design, as well as fork, borrow, and remix governance policy code from other communities. 
In support of these goals, the PolicyKit software infrastructure contains the following components to enable governance building: 
1) a software library to help users articulate policies in code, 
2) a continually-running server process that executes policies against actions in the community, 
3) a platform integration allowing PolicyKit to know when actions have been attempted on the community's platform and be able to execute passed actions on the platform, and 
4) a website where members can propose actions to alter the governance model and also author policies in a code editor.

In the following sections, we describe PolicyKit's abstractions, how to write policies in PolicyKit, and example implementations of a range of policies. These policies include democratic approaches such as a random jury deliberation and an election, as well as one inspired by Wikipedia's Request for Adminship (RfA) process. We also demonstrate a caucus involving multiple stages, a policy that calls an external API, and one that keeps track of reputation to unlock privileges.
Finally, we describe how to integrate a platform into PolicyKit and present implementations of integrations with Slack, Reddit, and Discord.
Our aim with PolicyKit is to lay the foundation for a broad space of interactive tools that empower communities to develop mature governance to suit their values and needs, and that in turn, give everyday people greater self-determination over the social platforms that influence their lives.


\section{Background and Motivation}


Despite early visions of the social web as an open and participatory space~\cite{barlow1996Declaration}, the first online communities were governed as technocratic autocracies, primarily due to the need for an ``admin'' with technical skills to own and operate the server that the community software ran on~\cite{schneider-admins2019}. 
While admins sometimes chose to carry out governance that was closer to anarchy or democracy in practice~\cite{dibbell1994rape,mnookin1996virtual}, admins still had the ultimate authority to shut down the server or kick users out.
Admins also had the ability to appoint users to ``mod'' roles in order to spread the work of governing. 

Undeterred by the need for admins, some communities still experimented with alternative governance models. A notable example is LambdaMOO, which moved from a benevolent dictatorship governed by ``wizards'' toward a petition system involving voting, where wizards were relegated to implementing the outcome of votes~\cite{mnookin1996virtual}.
While some custom code facilitated making a petition, most of the procedure of the petition system was carried out manually.
Some parallels can be drawn to English Wikipedia, another community with a more democratic model.
Like other peer production communities such as GNU/Linux, a core tenet of Wikipedia is its openness in permitting contributions~\cite{forte2009decentralization}.
However, conflicts arise, and over time, Wikipedia has grown a number of processes involving petitions and votes to address conflicts~\cite{im2018deliberation}. 
Similar to LambdaMOO, most of these processes are carried out manually, with the help of custom bots to perform some tasks such as documentation~\cite{muller2013work}, and final execution is often left to admins who can implement procedure outcomes~\cite{konieczny2009governance}.

Today, communities with community-created procedures for governance,
like LambdaMOO, Wikipedia, and open source projects like Debian, 
are the exception rather than the rule. This is because carrying out governance manually places a heavy burden on a small number of people. 
The individuals who have the power to implement the policies burn out due to the amount of work and stress involved in governing, sometimes leading them to leave the community entirely~\cite{konieczny2018volunteer} despite being core contributors~\cite{bryant2005becoming}.
Part of the issue is that by continuing to rely on users with privileged access, individuals ostensibly tasked with only policy execution still find themselves pressured and politicized by community members~\cite{im2018deliberation}.


As a result, most communities follow the default governance pattern made easy to adopt by their platform software. 
A few platforms have taken the step to build new governance models directly into their software. For instance, Slashdot and StackOverflow have a reputation system where users gain points in order to unlock greater power~\cite{lampe2004slash,mamykina2011design}. Other examples include League of Legends~\cite{kou2014governance} and Weibo~\cite{kou2017managing}, two platforms that created jury systems for adjudicating user conflicts.
However, these alternative governance models were designed and implemented by platform developers, as opposed to by  members of the community.
If community members wish to, say, change how reputation is calculated, they have no procedure for doing so, except to directly petition the platform in a form of collective action~\cite{centivany2016popcorn,matias2016going} or leave for another platform~\cite{fish2011birds,hirschman1970exit}. While the threat of such actions can blunt centralized power, resulting in ``benevolent dictatorships''~\cite{raymond1998homesteading}, the lack of structured ways to enact change can itself mask power~\cite{freeman1972tyranny}.



While unable to alter platform software directly, community members can still make use of software tools to support some aspects of governing, such as tools for deciding and carrying out policy.
For instance, many third-party tools exist to help communities come to consensus on decisions. These include tools to poll opinions that make use of visualization to surface points of agreement~\cite{faridani2010opinion,kriplean2012supporting}, as well as deliberative tools to help members consider each other's perspectives~\cite{farina2013democratic,klein2011harvest,kriplean2012you,zhang2017wikum}.
They also include systems to support delegating votes at scale~\cite{hardt2015google}, and systems that incorporate competition~\cite{malone2007harnessing} and cooperation~\cite{salehi2015we}.
While these tools provide novel ways for community members to give input on individual decisions, the question of how to provide and combine input is only one small part of a governance procedure.
A fuller accounting of governance must specify aspects such as who can provide input, when can they provide input, what happens after a decision is made, and how decisions can be overridden or vetoed.
In addition, none of these tools automatically interface with the community, and instead require an admin to carry out decisions.
Even in-house tools for collecting opinions such as Facebook Polls have no enforcement capability.

Other software tools are focused on better execution of governance but provide no mechanism for proposing or deciding on policies. Examples include Reddit's AutoModerator tool for mods to author rules for their subreddit~\cite{jhaver2019human}, or the wide variety of bots on Wikipedia that perform administrative tasks such as tagging, archiving, and fixing~\cite{geiger2010work,zheng2019roles}. 
In many cases, these automation tools are not, or do not start out, embedded in a platform but grow separately as one-off bespoke pieces of software~\cite{geiger2014bots}, making them more difficult to author and manage.
Additional tools exist that help mods create better policies~\cite{chandrasekharan2019crossmod,matias2018civilservant} but still are focused on the role of mods. 
Finally, some tools exist that are aimed at regular members of a community~\cite{geiger2016bot,jhaver2018online,mahar2018squadbox}, but these are primarily only used to enact policies for an individual as opposed to a group.



While all of these software tools help communities with one or a few aspects of governance, none are broad enough to describe and implement an end-to-end governance procedure of substantial complexity. PolicyKit's strength lies in its framework that structures and simplifies the task of writing software to support governance, but still provides the flexibility and power to implement a wide range of governance models.




\section{PolicyKit: Building Governance with Software}

We introduce PolicyKit, 
a software infrastructure for online communities to build governance that can be directly enacted on their home platform. 
Instead of articulating permissions, PolicyKit enables communities to author policies for how actions can be carried out, opening up a wider range of governance models, including more democratic ones.
PolicyKit provides a framework built on a set of software abstractions for succinctly articulating governance in code; this then allows governance to be carried out by software on the platform instead of manually by users.

\subsection{Abstractions: Actions and Policies}
Governance models such as a random jury or a direct democracy require input by one or more members before an activity can be approved. Other governance models may go through a series of checks in different stages.
In general, these governance models all require the articulation of some sort of  \textit{procedure} to arrive at a decision rather than a single permissions check.
PolicyKit's design is motivated by the insight that across these myriad procedures, we can describe the specific behavior that is being proposed separately from the rules being used to determine whether that behavior is allowed to proceed.
This insight draws from the work of political scientist and Nobel Laureate Elinor Ostrom, who studied offline communities governing common pool resources.
The IAD framework developed by researchers at the Ostrom Workshop broadly describes complex governance arrangements and centers around actors engaging in an ``action situation'' where they perform actions in light of an existing structure of rules~\cite{mcginnis2011introduction}.



\begin{figure}[tb]
  \centering
  \includegraphics[width=\columnwidth]{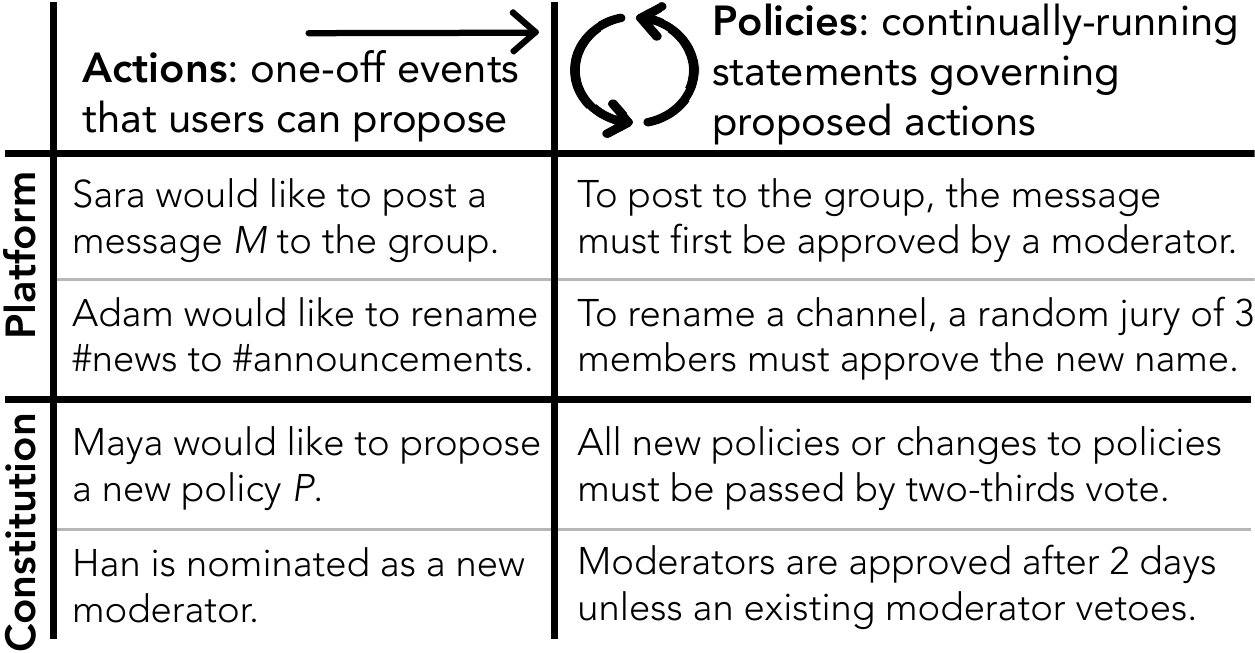}
  \caption{Left: \textit{actions}, or one-off events that users can propose. Right: \textit{policies}, or procedures that govern the actions on the left. Policies are stated here in natural language but in PolicyKit are written in code.
  We further differentiate actions and policies that happen on a \textit{platform} from actions and policies that relate to the \textit{constitution} of the community, which articulates procedures for creating and editing policies.}~\label{fig:table}
\end{figure}

Given this, the two main abstractions within PolicyKit are \textit{actions} and \textit{policies}. 
An action is a one-time event that can occur within a community and is typically first proposed by a community member.
In contrast, a policy is a declaration that must always be true and that governs some user capability. 
For instance, a policy for joining a community might be: \textit{``To join the community, a user must be approved by at least one existing member of the community.''}
We state policies here in natural language for ease of explanation; in a later section, we describe how PolicyKit expresses these policies in code.
An example of an action that would be governed by that policy would be, \textit{``Sanjay joins the community.''}
Policies can govern one or more actions. Thus, when a user attempts an action, before it can be carried out, any policies that govern the action must first approve it.
Additional examples of actions and policies governing them are provided in Figure~\ref{fig:table}, such as being added as a moderator or posting a message to a protected channel.

\subsubsection{Layers: Platform and Constitution}


Ostrom further distinguishes between several ``arenas of choice'' where action situations can occur. 
In the IAD framework, researchers separate out a ``constitutional choice'' layer for participatory change in a government's overall design~\cite{frey2019place, kiser1982three,mcginnis2014social}. This layer is separate from the others that are designated for the execution of that design.
Ostrom stressed the importance of a constitutional layer in governance, finding that successful communities follow the principle that  \textit{``...those affected by the rules can participate in modifying the rules''}~\cite{ostrom}.
Taking inspiration from these layers, we separate out the everyday actions that take place in a community and the policies that govern them  (the ``operational'' and ``collective choice'' layers of the IAD framework) from actions that are intended to change the governance model itself and policies governing those (the ``constitutional choice'' layer).
We coin these two layers \textit{platform} and \textit{constitution}, respectively, where this axis defines what the action or policy is targeting.
Examples are shown in Figure~\ref{fig:table}.

\textit{Platform} actions are one-time events that correspond to a user capability on the platform. This can include posting a message, joining a channel, or editing a wiki, depending on the platform in question. Platform actions tend to happen frequently as they make up the day-to-day activities of the community. 
Platform policies describe procedures that govern platform actions.
For instance, a platform policy could be that \textit{``All posts to this community must not use  swear words.''} A platform action such as ``\textit{Rosa would like to post }`hi!'\textit{ to the community''} would need to check to ensure the policy is met before executing.
\textit{Constitution} actions involve one-time events that \textit{alter} how governance is done, for instance, appointing a person to a new role, or changing an existing policy.
Constitution policies describe procedures that govern constitution actions.
For instance, a constitution policy could decree that \textit{``Any change to an existing policy must be passed by majority vote,''} and a constitution action could be that \textit{``Ayeesha wants to change the policy about majority vote to also stipulate that there must be a quorum of at least 10 voters.''}


\begin{figure*}
  \centering
  \includegraphics[width=1.9\columnwidth]{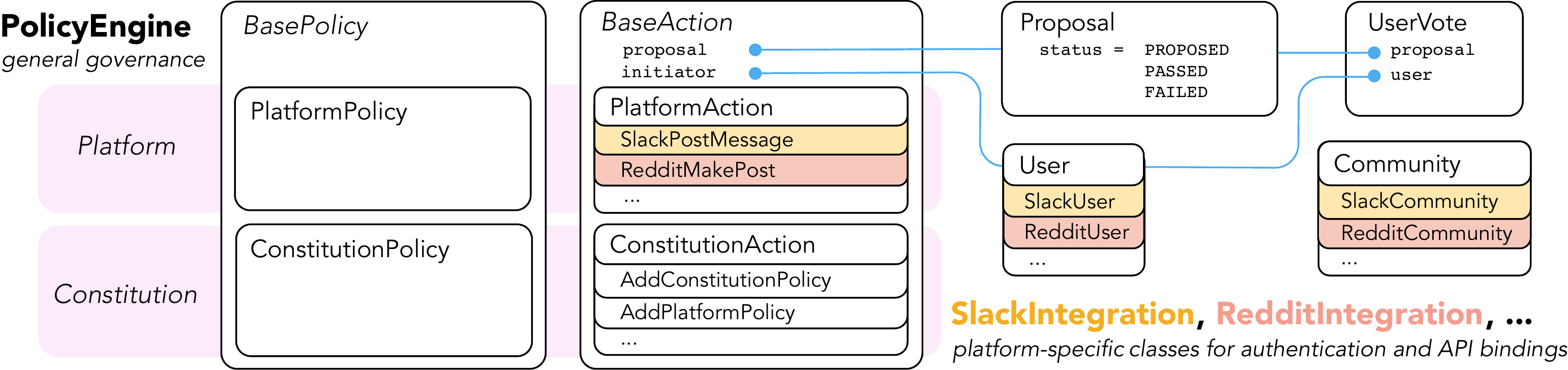}
  \caption{The PolicyKit data model contains general-purpose objects in the \texttt{PolicyEngine} library, as well as platform-specific objects in a platform integration library. The \texttt{PolicyEngine} library defines policies at the platform and constitution layer, as well as actions at the constitution layer. Each platform integration library defines platform-specific actions and their bindings to the platform's web API, plus a \texttt{Community} and \texttt{User} subclass.}~\label{fig:classes}
\end{figure*}

\subsection{PolicyKit Software Infrastructure}

We build off of these abstractions to provide a software framework for authoring policies, proposing actions, checking actions against policies, and carrying out passed actions on a community's platform of choice.
In order to provide these capabilities, PolicyKit consists of the following components:
\setlist{nolistsep}
\begin{itemize}
\itemsep0em 
    \item A \textbf{software library} for concisely authoring policies in the Python programming language,
    \item A \textbf{policy engine}, or a continually-running server process that waits for proposed actions and checks them against policies to see whether they can pass,
    \item A \textbf{platform integration} to a community's platform where members can attempt platform actions as well as vote on actions while going about their regular activities, and where passed platform actions can execute, and,
    \item A \textbf{web interface} where members can install PolicyKit to their platform, propose constitution actions, and author policies in a code editor.
\end{itemize}

Combining all these components, a community that wishes to use PolicyKit goes through the following process.
First, a platform integration must already exist for the community's platform; once any developer has written an integration, every community on that platform has the ability to install PolicyKit. We have thus far implemented integrations for Slack and Reddit.
To install PolicyKit to their community, a current admin of the community must go to the PolicyKit website and grant permission to the tool to be able to perform admin activities via the platform's API. 
Then, PolicyKit instantiates a new instance tied to the community and installs an initial governance model consisting of a single constitution policy governing all constitution actions. Our governance ``starter kit'' requires a direct majority rule for constitution actions to pass; in the future, additional starting policy positions could be provided for communities to choose between.

From there, community members can propose actions, as shown in Figure~\ref{fig:concept}. For instance, they could use PolicyKit's web code editor and the software library to author a new policy that replaces the initial constitution policy or creates the first platform policy---these proposed changes form constitution actions.
Meanwhile, the policy engine on the PolicyKit server is continually checking all proposed actions against existing policies to see if they can execute. 
When constitution actions pass, they execute on the PolicyKit server.

On the platform, members continue to conduct their everyday activities while PolicyKit's platform integration listens for actions. When an action is attempted, PolicyKit immediately receives an event from the integration and sends it to the policy engine to see whether the action can execute given existing policies. 
If it is not yet allowed to execute, the platform integration immediately reverts the action on the platform, notifies relevant users about votes they need to make, and listens for their votes.
Once the policy engine determines that a platform action can execute, it uses the platform integration to carry out the action on the platform. 

In the next sections, we describe details of the PolicyKit infrastructure. Following that, we describe extensions to the basic PolicyKit infrastructure as well as examples of policies authored in PolicyKit.

\section{PolicyKit Data Model}


The data model underlying PolicyKit, shown in Figure~\ref{fig:classes}, builds on the abstractions of actions and policies and is
comprised of the \texttt{PolicyEngine} library, which defines general-purpose classes related to governance,
as well as additional libraries, one for each platform to which PolicyKit can connect.
Policy classes contain fields that store user-written code for carrying out a policy, while 
action classes contain fields and methods to help carry out a particular action.
Within platform integration libraries, additional actions are defined that correspond to all governable actions on a platform.
For instance, our implemented \texttt{RedditIntegration} library defines a platform action \texttt{RedditMakePost} for making a post on Reddit using their web API.
As mentioned, actions are proposed by a community member and then must go through a procedure to determine if it can be executed.
Unlike permissions that can be evaluated instantly, procedures may take some time to complete, particularly if they are waiting on user input. 
To capture this information, each action class stores an \texttt{initiator} as well as a \texttt{proposal} object that stores the status of the proposal.
We also collect user input in the \texttt{UserVote} class.
For instance, if a policy states that an action can execute if a majority of the community has voted in favor, then part of the policy's procedure will check the proportion of votes in favor of that proposal.

\section{Policies and the PolicyEngine Workflow}

We now describe how to write and execute policies. In translating our abstractions into a concrete software design, we focus on the following design goals:
\begin{itemize}
\itemsep0em
    \item \textbf{Author concise, modular, composeable policies}: Policy authors should be able to express a broad range of governance models without needing to write excessive code. Logic that needs to be run repetitively to execute a policy need only be written once. Policies should be able to compose, reuse, and build upon logic from other policies.
    \item \textbf{Allow for human interaction and input}: Governance requires contention, disagreement, and relational work. Systems without levers for these behaviors will be abandoned~\cite{grudin1994groupware}. While policy authors should be able to create fully automated procedures, they should also be able to create procedures that give space to human relational labor, debate, and nuanced judgment~\cite{alkhatib2019street,mahar2018squadbox,salehi2015we}.
    \item \textbf{Minimize security risks to communities}: Communities should be able to view their policy code, audit policy decisions, test, and recover from undesired policy behavior.
\end{itemize}
Given these goals, we break down a policy into a set of functions that together articulate the actions that the policy applies to, how to determine if an action in its jurisdiction will pass, and what to do if an action passes or fails.
Functions allow for modularity, where each function is called at different periods in a workflow, with some being called repeatedly.

In Figure~\ref{fig:code}, we present an example of a platform policy that governs how channel names get changed on Slack. It stipulates that a random jury of three users in the community must approve any channel name changes by a majority vote within two days.
In Figure~\ref{fig:workflow}, we show how policy functions are called within the \texttt{PolicyEngine} workflow, a server process that periodically iterates through proposed actions. 
Once any new action is proposed by a user, either by invoking it on the PolicyKit website or on a particular community platform, it passes through the \texttt{PolicyEngine} workflow, first calling the \texttt{filter()} function for each policy.
All functions are passed the action object that is being evaluated and the policy object of the function.

\begin{figure}
  \centering
  \includegraphics[width=\columnwidth]{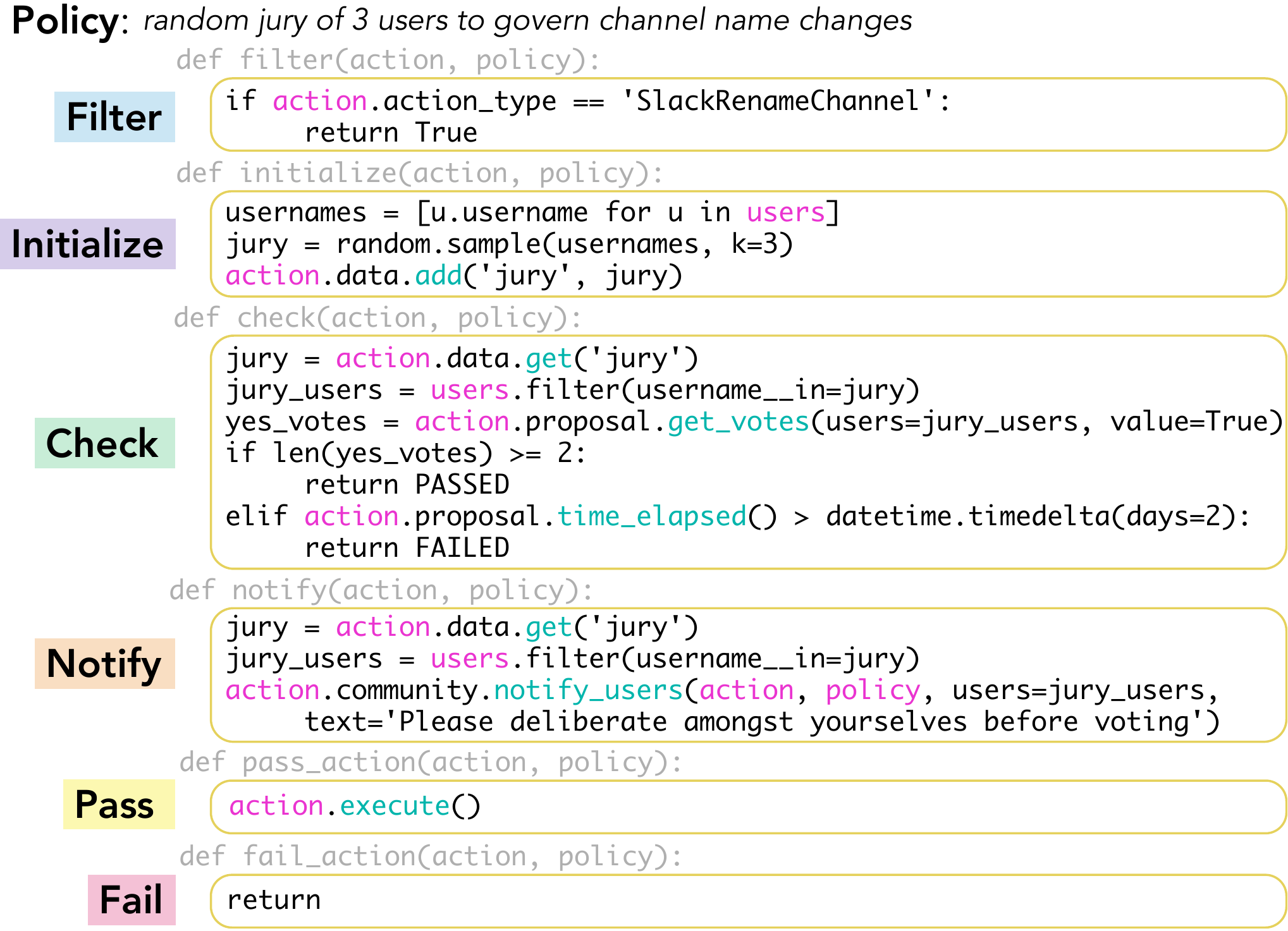}
  \caption{This \texttt{PlatformPolicy} governs channel name changes on Slack using a randomly selected jury of three users. Six functions must be implemented to create a policy. In their code, authors have access to community-specific objects (in pink) and helper methods (in cyan).}~\label{fig:code}
\end{figure}

\textbf{Filter}: This function specifies the \textit{scope} of the policy, or what types of actions the policy governs. 
The function returns \texttt{True} if the policy governs the action object passed in as an argument. For instance, if the policy is meant to cover only one type of action, the function can check the \texttt{action\_type} field of the action object. The policy could also filter on the initiator of the action or even the time of day, if, for example, the community has decided that Friday evenings are a free-for-all in a particular channel. 

\textbf{Initialize}: If \texttt{filter()} returns true, the action in question is considered in scope for this policy, and we move on to \texttt{initialize()}. Within this function, the author can specify any code that must be completed once at the start of the policy to set it up. For instance, in the jury example, \texttt{initialize()} selects the random jury who will decide on the action.

\textbf{Check}: The \texttt{check()} function specifies the conditions that need to be met so that an action has passed or failed. For example, it may test whether a vote has reached a quorum, or whether an elected individual has responded to the proposal. 
When created, all actions have a status of \texttt{PROPOSED}. 
New actions first encounter \texttt{check()} immediately after \texttt{initialize()}; this is so that in case the policy can already pass or fail, we can exit the workflow early. 
For instance, if there was a policy that holds messages containing profanity for review by a moderator, the policy would automatically pass actions that do not contain profanity.
As long as an action is still \texttt{PROPOSED}, \texttt{check()} will run periodically until it returns \texttt{PASSED} or \texttt{FAILED}. 
If \texttt{check()} does not return anything, \texttt{PROPOSED} is presumed.
For instance, if a policy calls for a vote from users, such as in the case of the jury policy in Figure~\ref{fig:code}, it may take time for the required number of votes to come in. 
The policy's \texttt{check()} function could also specify a maximum amount of time, at which point the action fails; in the jury example, the action has a maximum of two days.

\begin{figure}[tb]
  \centering
  \includegraphics[width=\columnwidth]{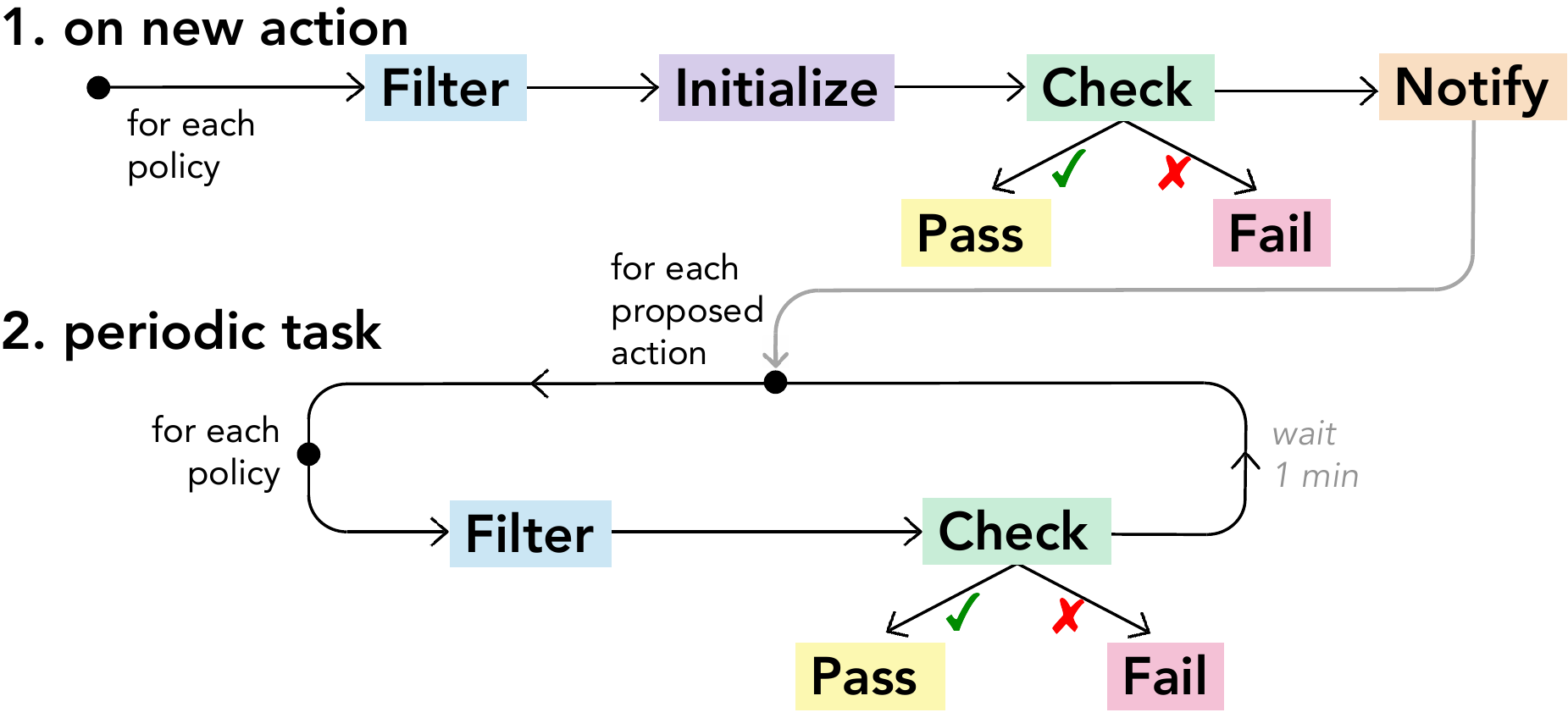}
  \caption{The workflow in \texttt{PolicyEngine} to check actions against policies:
  Once an action is created, we call \texttt{filter()} for each policy to determine if the action is in scope. If it returns \texttt{True}, \texttt{initialize()}  runs to set up the proceedings (e.g., choose a jury),  followed by \texttt{check()} to test if the action can pass already. If the action fails or passes at this point, we can exit the workflow. Otherwise, we run \texttt{notify()} and move into a periodic task where all actions that still have \texttt{status=PROPOSED} are periodically checked against all relevant policies until they pass or fail.}~\label{fig:workflow}
\end{figure}

\textbf{Notify}: If the policy involves reaching out to one or more community members for input, then the code for notifying members occurs in this function. 
While policy authors can send messages to users in any function, this function is specifically for notifications soliciting user input.
Authors may use the helper method \texttt{notify\_users()} to send messages to community members, with ability to customize the post. For instance, the notification post can include instructions, such as to deliberate the action before voting.
This function is only run once, after a new action does not return \texttt{PASSED} or \texttt{FAILED} from the first \texttt{check()}, so as to not unnecessarily notify users.

\textbf{Pass}: This function runs if an action is passed. Each action class implements an \texttt{execute()} method that policy authors can call to carry out the action. Other code that could go here include post-action clean-up tasks such as announcing the outcome to the voters or to the community.

\textbf{Fail}: This function runs if the action fails to pass the policy. For instance, the author could add code to invoke fall-back actions due to failure, or share the outcome privately with the proposer alongside an explanation of why the action failed.

\subsection{Software Library and Security}
Within each policy function, users have access to all community-specific objects such as users, policies, and actions, along with their public methods, in addition to the direct policy and action in question. This is shown in Figure~\ref{fig:code} with the objects highlighted in pink.
Filtering can be done using the Django framework's query syntax, as PolicyKit is an extension of the Django framework, and community-specific objects are passed in as a Django \texttt{QuerySet} as opposed to a list of objects.
Users can also access related objects using Django's object-relational mapping.
As seen in the cyan-colored text in Figure~\ref{fig:code}, users have access to a number of helper methods attached to objects to make common calculations such as counting votes or determining how much time has elapsed since a proposal.
The full API documentation of user-accessible classes, fields, and methods, as well as the open-sourced code, is available on the PolicyKit website: \url{https://policykit.org}.

\begin{figure*}
  \centering
  \includegraphics[width=2.1\columnwidth]{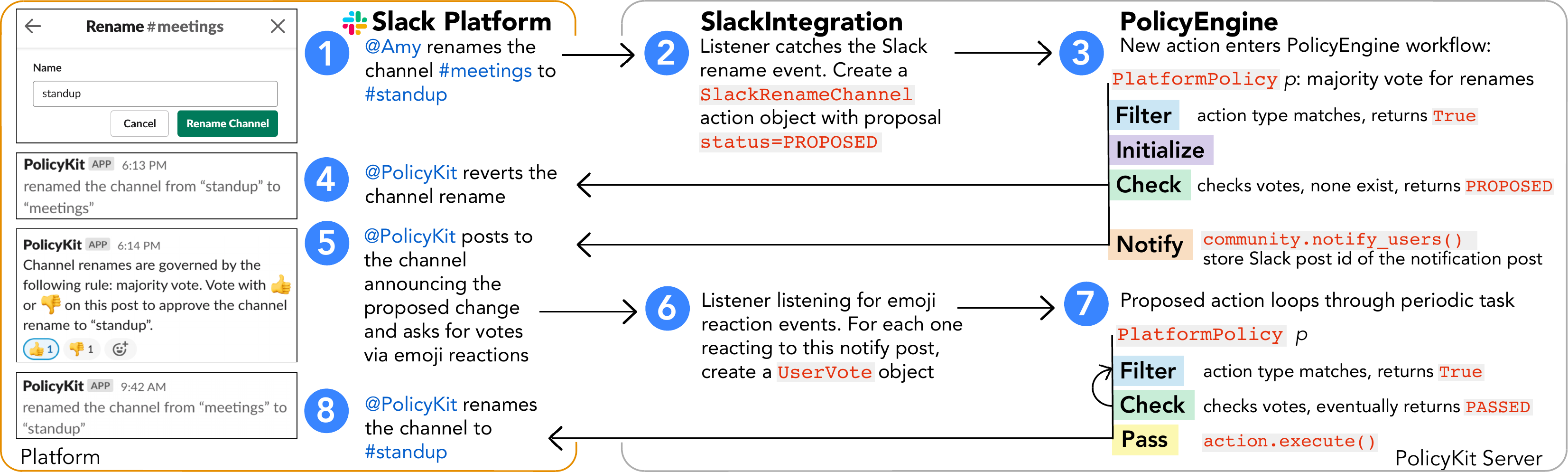}
  \caption{We show how PolicyKit integrates with platforms, using the example of Slack. 1) A user performs an action on Slack. 2) The \texttt{SlackIntegration} library running on the PolicyKit server has listeners for Slack actions. After catching an action, it creates a new \texttt{PlatformAction} object of that action type. 3) The proposed action enters the \texttt{PolicyEngine} workflow. If there is a policy that governs that action, it will run \texttt{check()} to see whether it can pass or fail. 4) It cannot, so the action is reverted on the platform. 5) The policy's \texttt{notify()} function runs, posting a notification to the Slack channel. 6) As votes are cast, a listener creates \texttt{UserVote} objects tied to the action. 7) The action is still looping through the policy workflow, where eventually it passes due to a change in votes. 8) Finally, the action executes on the platform after passing.}~\label{fig:engine}
\end{figure*}

To enable custom storage, each policy object and action object has a \texttt{data} field that points to a JSON object.
For instance, in the jury policy, a new random jury is selected for each action that gets proposed, so the jury members for that action must be stored in the action's \texttt{data} object. Records that could be stored in the policy's \texttt{data} object could include the failure rate of prior initiators (shown in the example below) or a list of recent jurors to avoid oversampling.

  \includegraphics[width=\columnwidth]{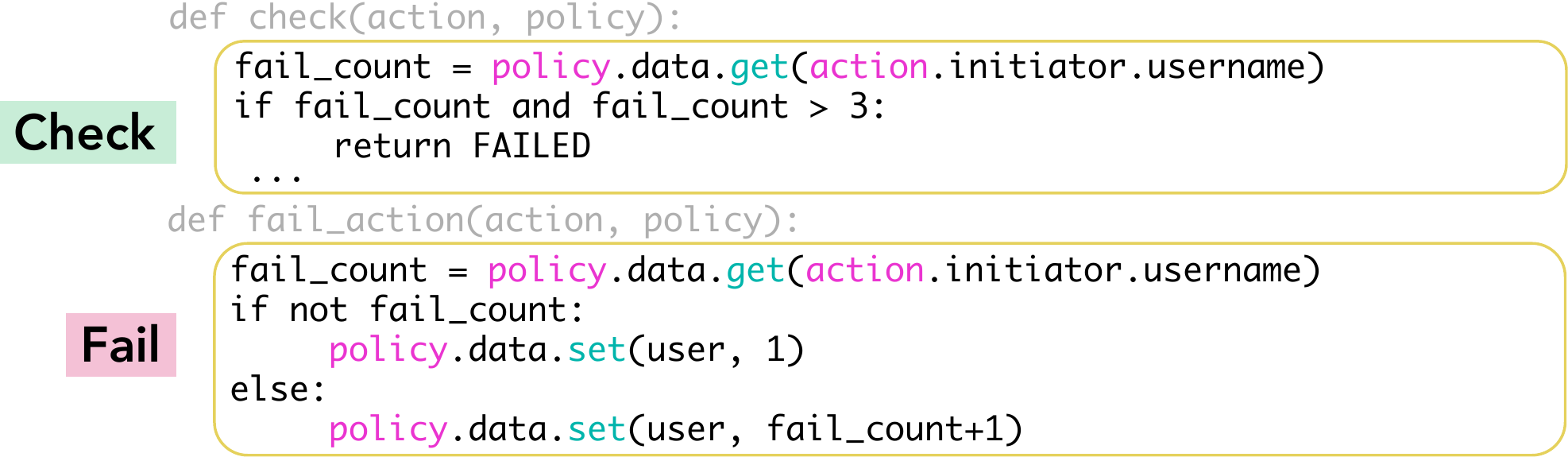}

Since PolicyKit policies can effect substantial changes to a community, we must consider security. We designed PolicyKit to ensure two principal security goals: (1)~incorrect policies could be audited and their actions reverted; and (2)~new policies could be field tested without risking the community. All PolicyKit policies are publicly visible to any community member on the PolicyKit website. To handle buggy policies, PolicyKit maintains a log of all actions, including their final disposition (e.g., passed, failed), and the policy that made that determination. This audit log allows the community to understand why an outcome occurred. All actions must be revertable, so if a policy passed an action in error, it can be undone through another action. To enable communities to field test new policies without risk, PolicyKit allows policy authors to leave the body of \texttt{pass()} and \texttt{fail()} empty---and in particular not to call the action's \texttt{execute()} method---allowing the community to use the audit log to track what a policy \textit{would have done} if it were live. After a trial period, community members can then activate the policy by passing a change that fully implements those functions.

PolicyKit's software architecture sandboxes code where appropriate. Policy authors cannot call private methods, access objects not associated with their community, or import libraries beyond a few that have been pre-imported. They also cannot directly query the database, though they can filter on the community-specific objects that they already have access to.
We reflect on other attack models (e.g., subtly malicious policy code, disgruntled admins pulling the plug on PolicyKit) in the Discussion section.


\section{Integrating Platforms with PolicyKit}

PolicyKit is an application that sits on its own server. However, it would be prohibitive if users of a social platform needed to go to PolicyKit for every governance task, such as proposing an action or voting on a proposal. In addition, as PolicyKit needs to enforce policies, it must have a way of stopping and allowing actions that are carried out on the platform itself, since the platform already has an existing governance that PolicyKit must supersede. 
These capabilities are defined in platform integration libraries that can be developed for any platform to connect with PolicyKit.
Once a single developer has created an integration using a platform's web API, any community on that platform can use PolicyKit.
Figure~\ref{fig:engine} presents an example of an action that is attempted on a platform before being governed by PolicyKit.
This demonstrates the following necessary components of a platform integration library as well as requirements of the platform API. 

In order to install PolicyKit to a community, there must be an \textbf{authentication workflow}, such as OAuth, for at least one admin or mod account to give access to PolicyKit so that it may govern a broad set of actions, including privileged ones. 
The platform integration must also specify ways to \textbf{send messages} to users on the platform. For example in Figure~\ref{fig:workflow}, we message users to solicit votes from the jury.
In order for PolicyKit to govern actions, it must know what \textbf{platform actions} are possible; these are specified via the creation of \texttt{PlatformAction} classes.
Actions typically are carried out via web API endpoints provided by the platform that are then made available through an \texttt{execute()} method in the action class and undoable via a \texttt{revert()} method. 
Finally, the integration must incorporate a \textbf{listener} to listen for user actions on the platform as well as a listener for votes on a notification message.
For instance, votes could be recorded via an emoji reaction or a reply to a notification message.




We have implemented platform integrations for the platforms Slack, Reddit, and Discord, with some differences between them due to their web API. 
For instance, our Slack and Discord integration collects votes via emoji reaction to the notification message, while the Reddit integration looks for replies such as ``+1'' or ``-1'' to the message. We cannot collect Reddit votes via their upvoting mechanism because their API does not expose votes per user.
We additionally investigate the feasibility of integrating other common community platforms, including Facebook Groups, Twitch, Github, and Discourse, by inspecting their API documentation. All inspected platforms have the necessary API components, including OAuth authentication, event listeners, notifying users, and action execution and revert, to be integrated with PolicyKit. Facebook Groups is the sole exception: no API endpoint exists to support reverts such as deleting a message. 




\subsection{Performance and Scaleability}
Once a community installs PolicyKit, it listens for all governable events that occur within the community on the platform.
If the platform API does not support real-time webhooks, there may be a slight lag in carrying out policy due to polling---however, this lag is negligible considering the speed at which actions typically happen in a discussion-based community. However, webhooks would be needed to integrate PolicyKit with, say, a gaming platform like Minecraft.
Otherwise, checking for actions and looping through policies are fast executions that do not cause appreciable load on a server. 
As larger communities use PolicyKit, we can move towards multiple servers; since actions are atomic, they can be considered in parallel.

\begin{figure}
  \centering
  \includegraphics[width=\columnwidth]{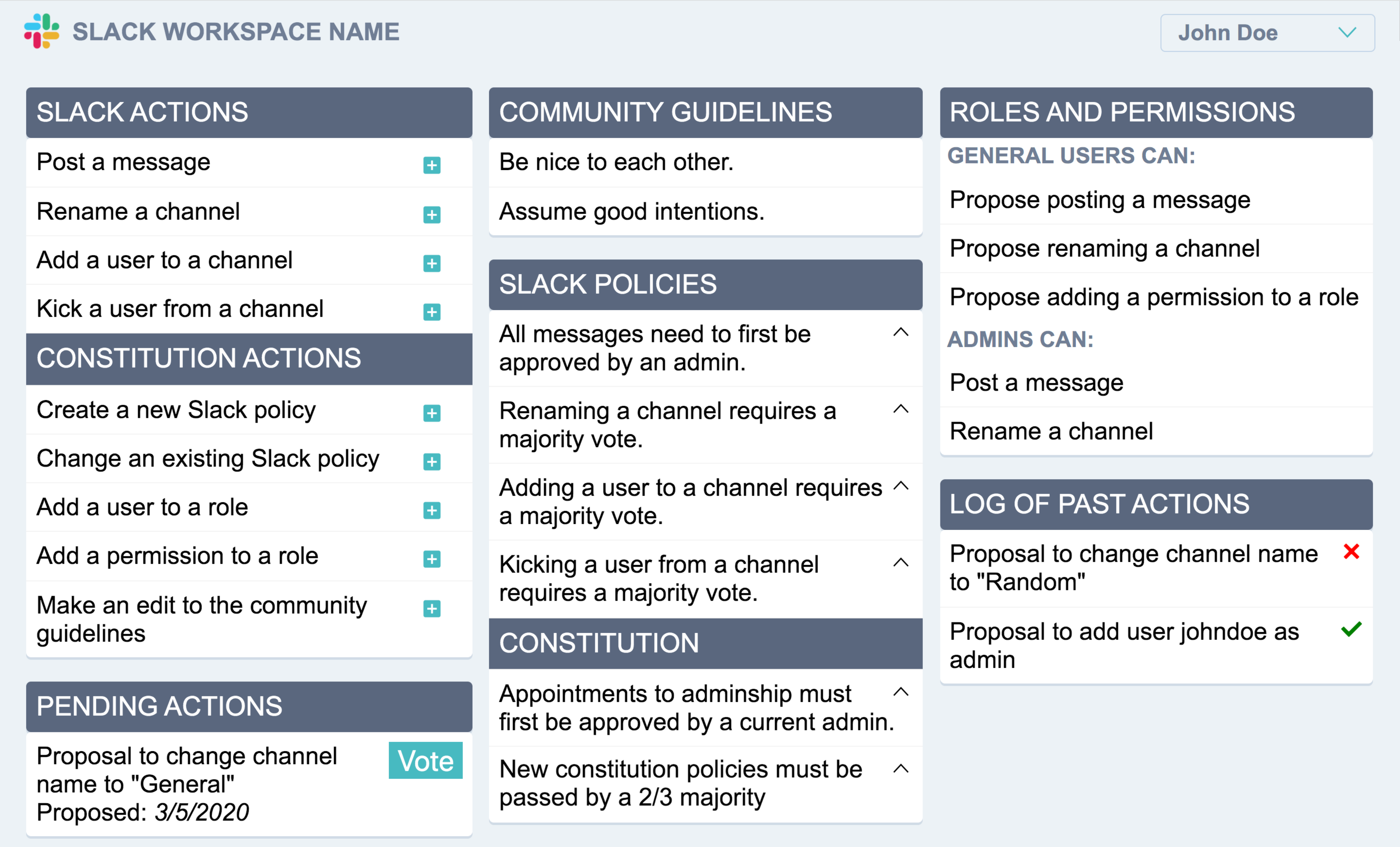}
  \caption{Screenshot of the PolicyKit homepage.}~\label{fig:screenshot}
\end{figure}

\section{Web Interface}

The PolicyKit web application allows community members to view their community's policies and audit log, as well as draft new policies. 
If desired, members of the community can propose any  action (left column of Figure~\ref{fig:screenshot}). 
Clicking on an action leads to a page for authoring that action proposal. For most actions, this simply involves filling out fields in a form. 
However, for the constitution actions for proposing a new policy or changing an existing policy, the authoring page opens to a code editor where users implement the functions that make up a policy. The editor includes syntax highlighting and code autocompletion along with library documentation to assist the author.
In the middle column, users can see the existing policies, including a natural language description of each policy written by the policy author. Clicking in to the policy allows users to inspect the code behind the policy.
On the right-hand side, users can see a log of past actions that have passed or failed and the policies that governed them.

\section{Data Model Extensions}

Previously, we described the basic data model focused on policies and actions. 
While this structure along with the ability to write open-ended code for policies allows for the creation of a wide range of policies, we also wanted to provide ways to more easily create certain types of policies that are common in many governance systems today~\cite{myers2000past}. 
We describe extensions to the basic PolicyKit data model to support these abilities.


\textbf{Roles and Permissions:} As mentioned, most platforms follow a permissions model. While it is possible to implement such a model within PolicyKit purely in policy code, since the pattern is common, we extend the data model to incorporate it.
The data model additionally contains a \texttt{Role} and \texttt{Permission} class and a series of constitution actions to alter roles.
A role contains a set of associated permissions and set of users who have that role.
Three \texttt{Permission} objects exist for each type of action. 
 Permission to \textit{view} an action type permits a user to view an action's log on the PolicyKit website. Permission to \textit{propose} an action type gives the ability to create action objects of that type, either via the PolicyKit website or by invoking the action on the platform. Permission to \textit{execute} an action type means that a user can perform an action regardless of any existing policies governing it.
Platform integrations can also define additional permissions.
Each community starts out with a \texttt{base\_role} comprised of all members.
As part of the current ``starter kit'', all users have \textit{view} and \textit{propose} permissions on all actions. Over time, these permissions can change as users propose changes or create new roles. In the future, a different starter kit could replicate the roles of admins and mods that exist in the community to allow for a more gradual transition to a new style of government.


\textbf{Documents for Codes of Conduct, Written Policies, etc.:}
Not all policies within a community can be expressed or enforced via code. For instance, some are general guidelines that users should keep in mind, such as ``\textit{Assume good intentions}''.
Many platforms have text space for community leaders to write these guidelines; for instance, every subreddit has a Rules section, as does Facebook Groups.
Since communities may desire to keep and also govern non-executable guidelines, we add a \texttt{Document} class to the data model, as well as constitution actions to alter documents. We start each community off with a single empty document. There is a rich-text editor for authoring documents on the PolicyKit website, and documents are also displayed on the homepage.

\textbf{Action Bundles:} 
In the current framework, actions are proposed and considered one by one. However, sometimes multiple actions need to be considered as a group of possible actions. A common use case is a vote between a selection of possible actions: for example, the community will elect either Pablo or Xinlan as the new President. 
To facilitate these governance events, we introduce an \texttt{ActionBundle} class. This class is an action just like other actions but it links to a bundle of other actions. Action bundles have two types: \textit{election} and \textit{combination}. When a user puts forth an \textit{election} action bundle, they are proposing to select from a set of action options. When a user puts forth a \textit{combination} action bundle, they are proposing that all the actions be considered together as one. Proposals are made via the PolicyKit website. As elections involve selecting from multiple options, platform integrations must handle them in their notification and vote listener implementation. For instance, our Slack integration
has a default template for notifying users about election options and listens for number emoji reactions as votes.

\textbf{Policy Bundles:} Similar to action bundles, sometimes multiple policies need to be considered together as a single multi-stage policy. For instance, it is common in offline governments, such as the U.S. federal government, for a policy to involve a vote by one set of people first; if a proposal passes that stage, it is then voted on by a different set of people, and so on.  
To permit these kinds of linked procedures, we incorporate a \texttt{PolicyBundle} class. Policy bundles are just like other policies but they link to a bundle of other policies. By grouping related policies together in this way, it is possible to propose a constitution action to instate or modify a policy in the same way for single-stage versus multi-stage policies.

\textbf{Datetime Triggers for Actions:} While we have only discussed actions triggered by users, it is also possible to programmatically create a new action within a policy's function. This can be useful in some cases; for instance, if an election fails because there was no majority, the policy can launch a new follow-up election.
However, sometimes actions should only become active after some period of time or on a certain date. To facilitate this, actions can be proposed along with a \texttt{datetime\_trigger} field that stores when the action becomes active.

\section{Examples}

We now present a series of examples to demonstrate the expressivity of the software library for authoring policies. Except where noted, the entirety of the code for these policies is included below.

\subsection{Wikipedia Request for Adminship}
After many years of evolution, Wikipedia has developed a process for promoting editors to admins~\cite{burke2008mopping}. Below, we demonstrate a constitution policy that takes inspiration from this Request for Adminship (RfA) process. In order to be appointed to the role of Admin, a user must have over 500 edits and 30 days of tenure. (We omit code for gathering that information due to space.) Their request is posted to an RfA noticeboard, where they receive votes and respond to questions. Then, after a period of 7 days, only a person with the role of Bureaucrat can approve their request. A Bureaucrat can also close the request before the 7-day period.

  \includegraphics[width=\columnwidth]{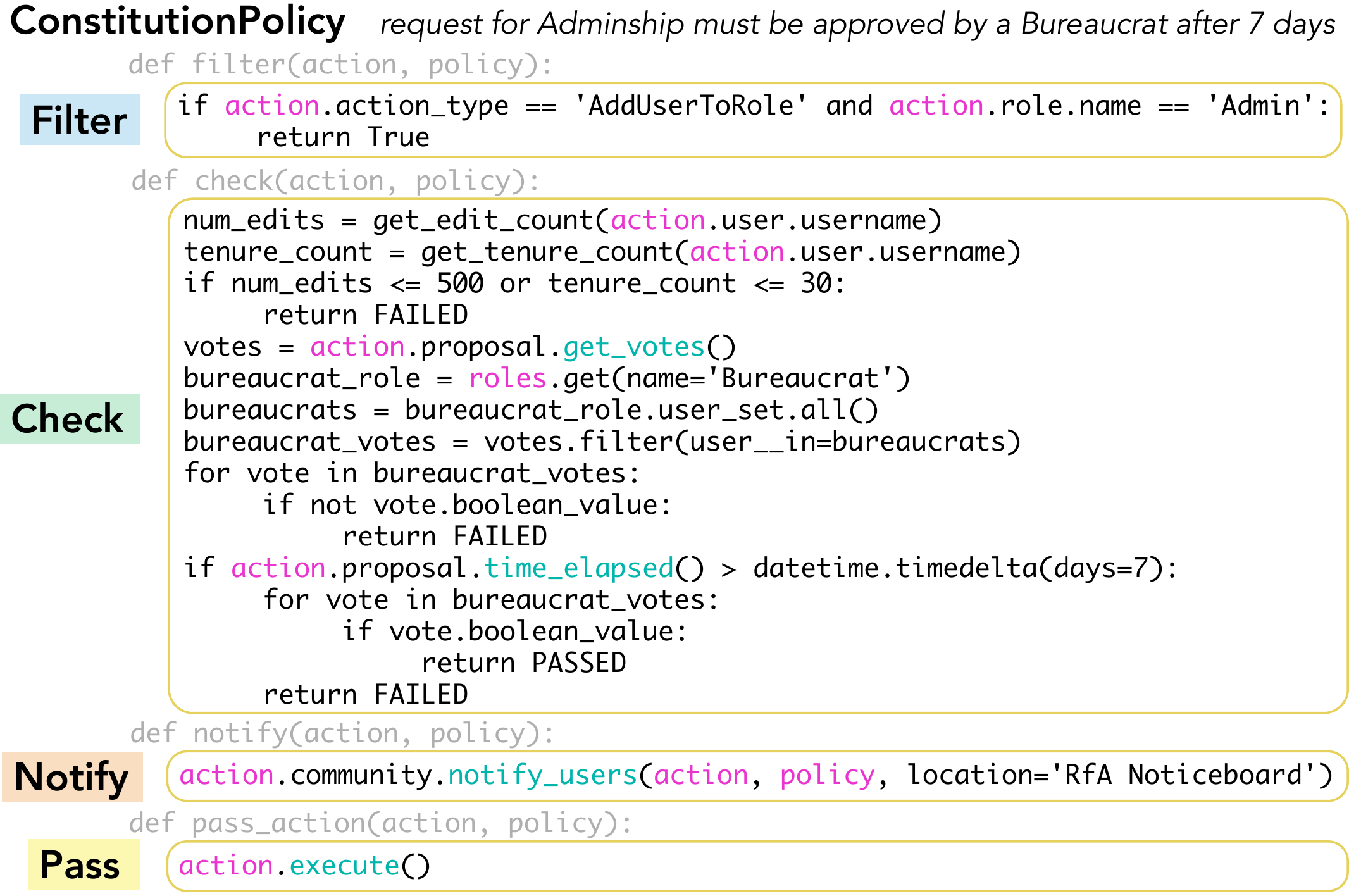}


\subsection{Election}
It is not uncommon for communities to hold regular elections to a position of leadership. 
Below is a constitution policy to govern elections to elect the next Steward of a community. The election runs for 5 days, after which all votes are tallied if there is a quorum of 25\% of members. In the \texttt{pass()} function, we determine which candidate has the most votes, counting maximum one vote per user. Then we remove the current Steward from the role and appoint the winner of the election by calling \texttt{execute()} on the specific action object in the bundle that adds that user to the steward role.


  \includegraphics[width=\columnwidth]{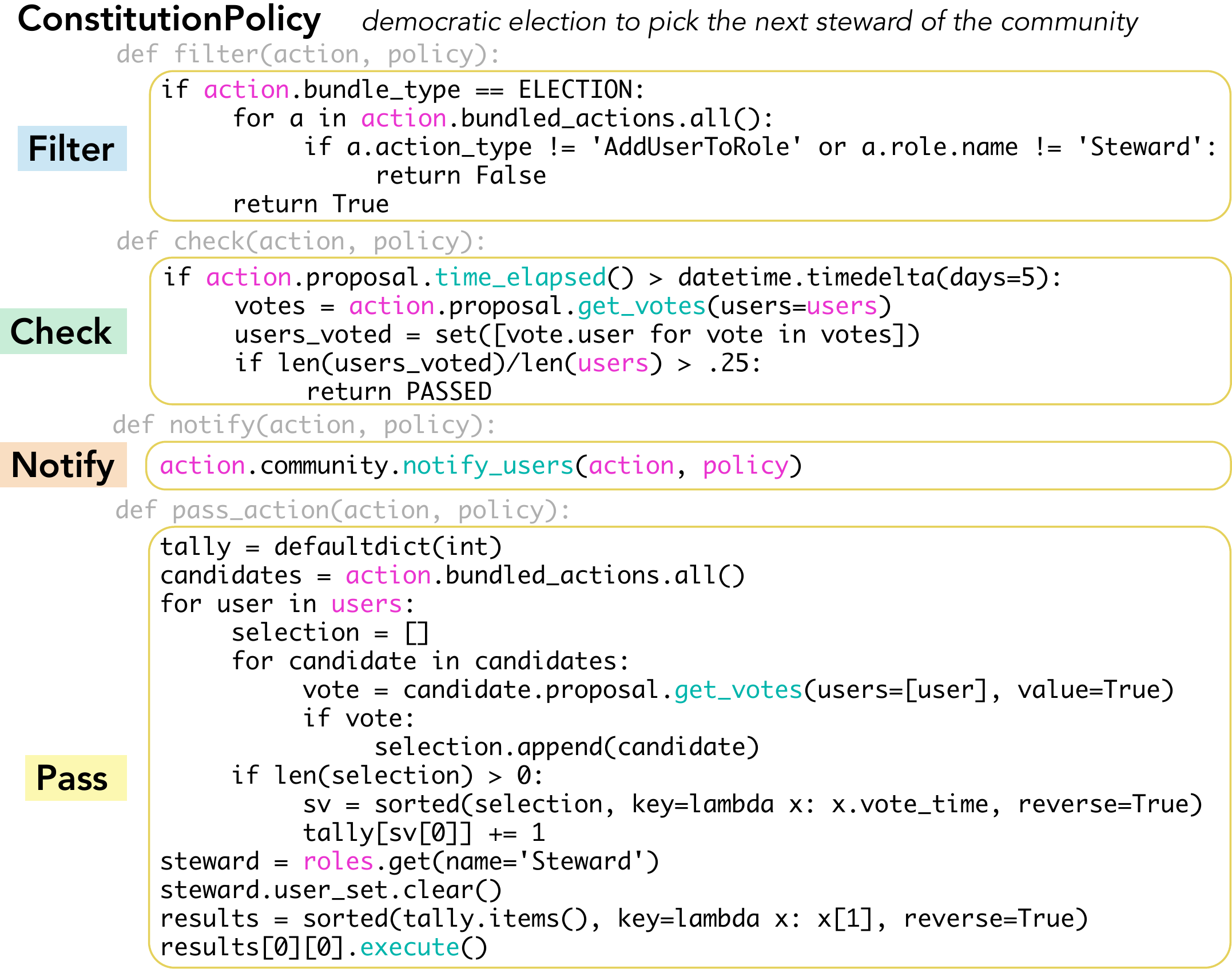}

\subsection{Two-Round Caucus: Pipelined policies}
Two-round caucuses are common in many U.S. state primary elections. 
Below is an example of a policy bundle that implements this procedure.
Unlike the election shown previously, candidates below a vote threshold are dropped after Round 1 and then users who voted for them get to switch their vote to another candidate in Round 2. 
We demonstrate how composition can be used across policies by calling functions from the prior election policy instead of duplicating its logic.
In both of the bundled policies, we use the action's \texttt{data} field within \texttt{filter()} to make sure the policies are executing in the right order. In Round 1's \texttt{pass()}, after tallying up votes, we remove non-viable candidates from the action bundle and remove votes for those candidates.
We also store the voters who can switch their votes inside \texttt{data} so that they can be notified in Round 2. 
We omit \texttt{check()} functions that would be the same as in the election example.

  \includegraphics[width=\columnwidth]{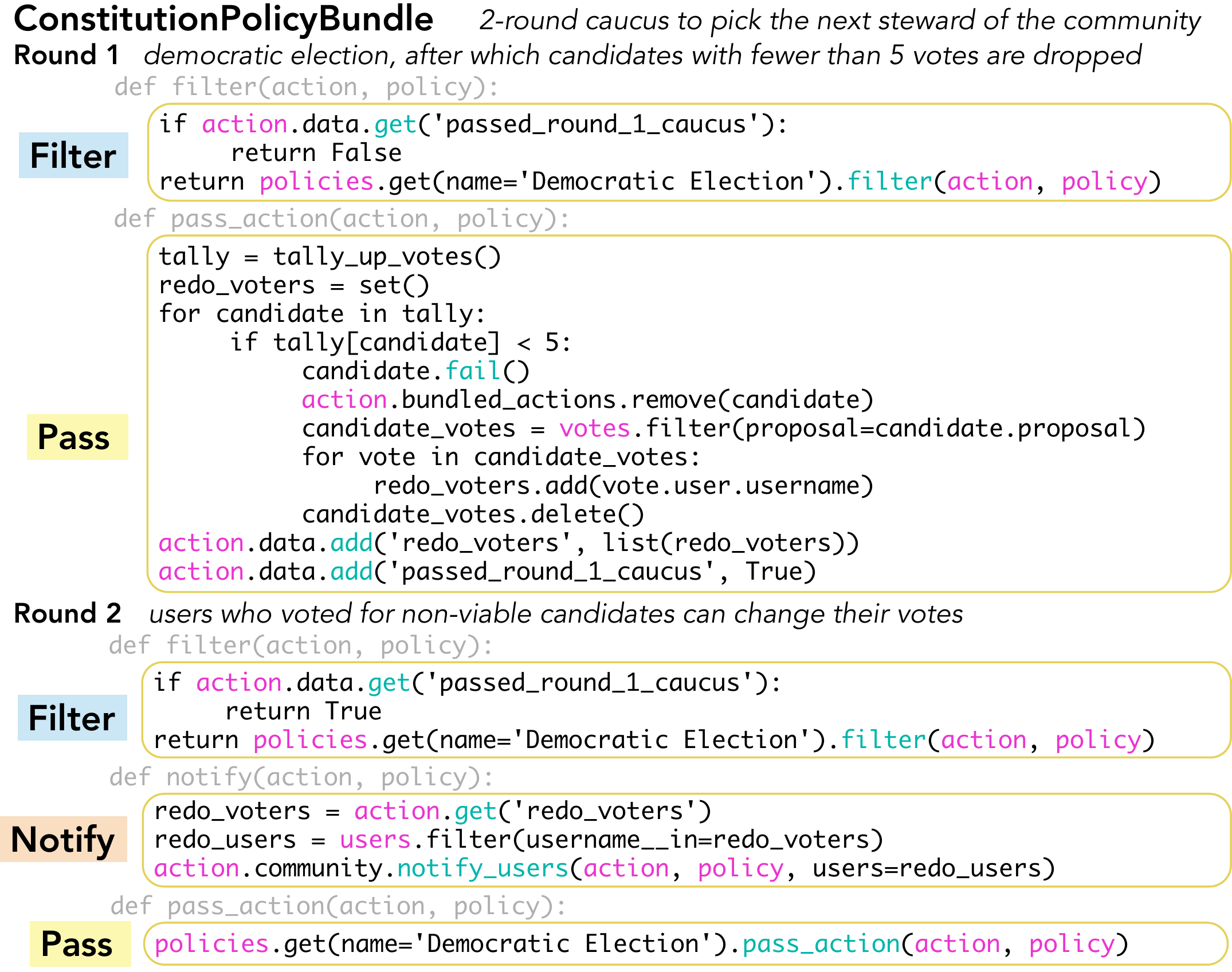}

\subsection{Toxicity Filter on Comments}

PolicyKit can integrate with external web APIs to support governance. In this example, a platform policy calls the Jigsaw Perspective API~\cite{perspective}, to return a toxicity score for the text, which the policy uses to filter out toxic comments. By being able to call external APIs, PolicyKit policies can use resources on the internet to augment their capabilities. Communities can also develop additional governance capabilities outside of PolicyKit, allowing policies to become arbitrarily complex and store data externally that is larger than what can be reasonably placed inside a JSON object. For instance, a policy that does not allow users to post links that have been posted to the community before needs to store a table of all prior links posted. If this list becomes long, it would be faster to store the links in an external database that a policy can then query.

  \includegraphics[width=\columnwidth]{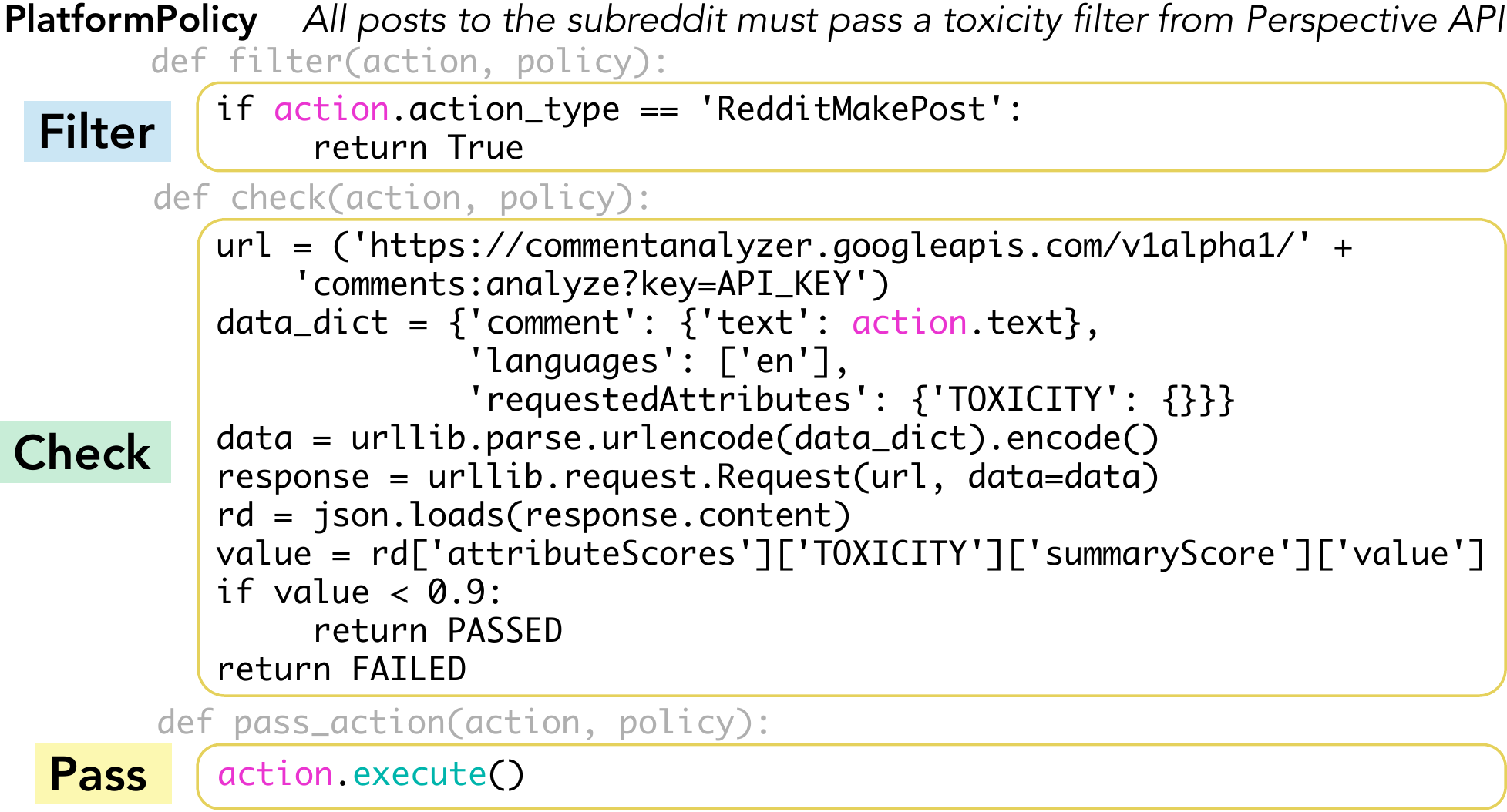}

\subsection{Reputation System}
Taking inspiration from sites like StackOverflow, PolicyKit allows reputation to be tracked via a policy that filters for relevant actions and updates the policy's \texttt{data} field to store user reputation. A separate policy can then check the reputation score and give users privileges based on that score.

  \includegraphics[width=\columnwidth]{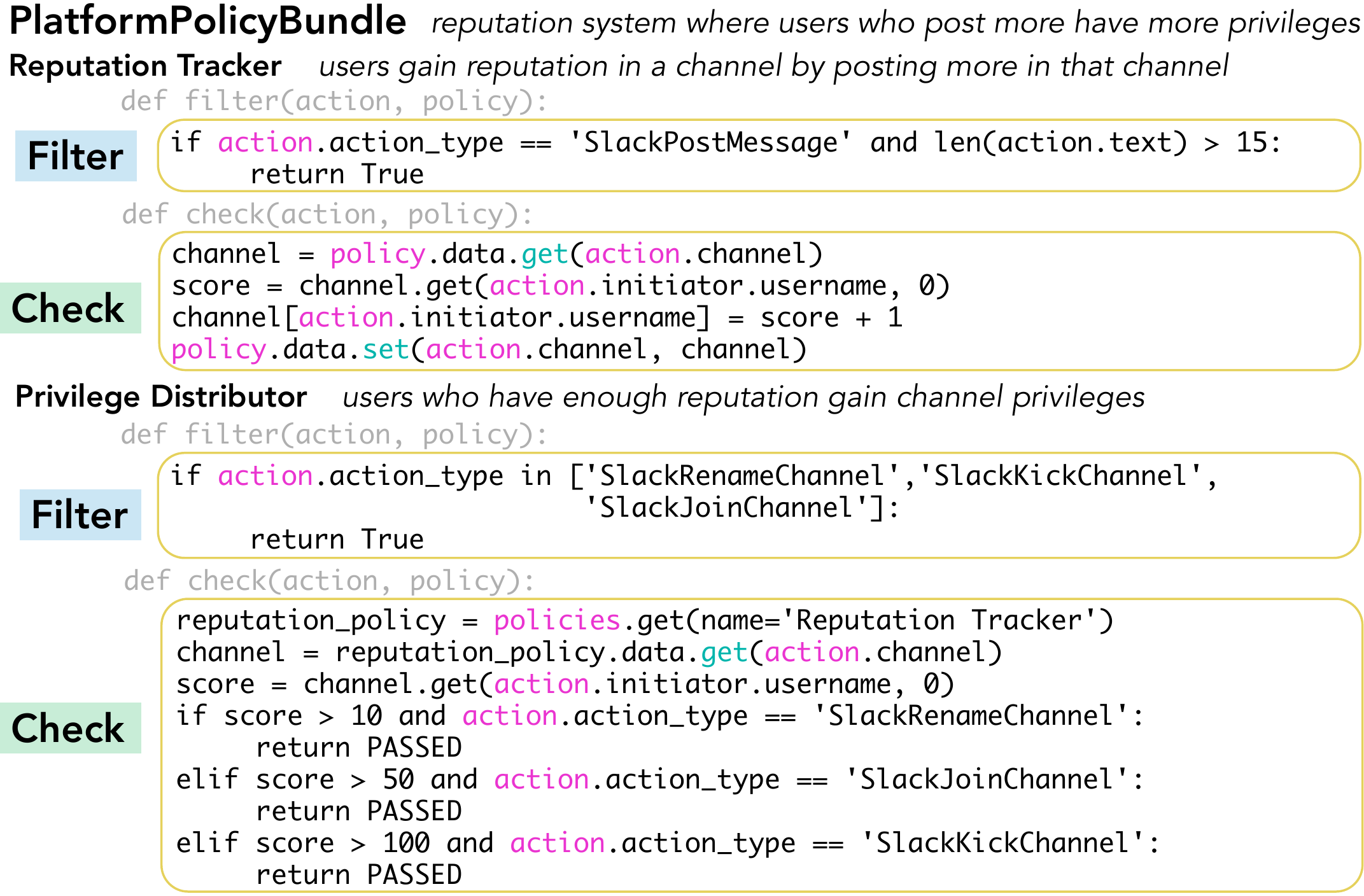}



\section{Discussion}

PolicyKit provides a low-level infrastructure for building governance in online communities. The introduction of PolicyKit unlocks a range of higher-level extensions that could be built on top of the system and enables new research directions.

\subsection{Extending Expressivity of Governance Authoring}

While there are many governance models that PolicyKit can implement, our framework could provide additional scaffolds to more easily express certain complex forms of governance.
For instance, while users can author multiple policies that govern the same action, the current process for determining what happens if they conflict is cumbersome---involving using policies' data store to set precedence as we showed in the caucus example. 
More broadly, an extension to the data model could support logic to define different kinds of relationships between policies beyond policy bundles. 
There may also be use cases for policies that dynamically activate or that activate based on modes. For instance, a school's online community might have one set of policies they wish to activate only during the first week of school when many new members are added.

In addition, there are no existing scaffolds to support collaboration on proposals or update pending proposals, in order to, for instance, merge two proposals together. To support this activity, the website would need additional features surrounding action proposals, such as comments or suggestions. There is also a question of the governance of the proposal itself. Currently, proposals have an initiator who would serve as the owner of the proposal. But in the future, proposals might be authored by multiple people or allow input from all members. A simple example is hosting an open election for a role, where anyone can nominate someone by adding to the action bundle.

\subsection{Building Towards End Users}

The current version of PolicyKit requires programming skills to author a policy. 
Additional tools such as a debugger, logging mechanisms, and tools for monitoring and testing could help make the programming process easier for users.
Communities that prefer not to code could also copy scripts from other communities, much like how Reddit mods pass around regular expressions that feed into Reddit Automoderator. 
However, requiring any programming even if simplified will likely deter some non-technical members from contributing and elevate the power of technical members.
For instance, while the code underlying policies provides a measure of transparency, in practice, end users may not be equipped to judge policies.

Our vision for PolicyKit is that eventually any community will be able to use it, regardless of programming skill.
To that end, this work serves as a low-level framework---a governance kernel---upon which higher-level systems can be built that better target end users.
Moving forward, PolicyKit could include templates for declaratively authoring common types of policies. For instance, a basic ``direct democracy'' form would simply need the author to specify
the minimum number of votes, the ratio of votes needed to pass, and the maximum amount of time to vote.
As communities use PolicyKit, we could study common policies to inform new templates and discover what configurations are important to expose in a template~\cite{juries}.
In addition, currently only one developer is needed to author a platform integration before anyone on the platform can use it. However, PolicyKit could include forms for end users to specify platform integrations. 
As platform have similar authentication workflows and API-calling logic~\cite{scrapir}, the main work is just to list all actions with their corresponding web API endpoints and URL parameters.




\subsection{Designing for Good Governance}

Using PolicyKit, communities can try out different governance models to see what suits them. 
Existing infrastructure for experimentation~\cite{matias2018civilservant} could even be incorporated to help communities trial new designs.
But coming up with good governance can be tricky.
Researchers have established principles based on studies of existing online communities or offline governments. For instance, research has found that people who get their post removed are less likely to re-offend if they receive a removal explanation~\cite{jhaver2019does}. 
Other research uncovered preferences for alternatives to typically punitive governance systems that instead center victims~\cite{schoenebeck2020drawing}.
However, community members likely do not know about all this prior research.
Currently, PolicyKit can be used to create governments with all manner of power distributions and procedures that are just or unjust, efficient or bureaucratic.
Instead of every community learning through trial and error, PolicyKit could help communities succeed faster by incorporating suggestions to authors based on prior literature.
A centralized repository of policies would also allow communities to more effectively learn from and build upon the work of others. 

Finally, PolicyKit could help communities better transition through different stages of their life cycle~\cite{ransbotham2011membership}. 
Currently, communities begin with a starter kit consisting of a blanket constitution policy requiring majority vote. 
More research is needed to determine what kinds of governance are best suited to new communities and what other options we could provide at the outset. 
Conversely, communities that are older and already have a strong set of norms and rules may need a different approach to migrate to use PolicyKit.
Communities may also need to quickly adapt their governance as they grow~\cite{kiene2016surviving}, as 
larger communities tend to require more rules~\cite{frey2019emergence}.
For example, as contributors grow in open-source projects, 
original owners shift to administrative roles, and the organizational structure changes to a distributed coordination model~\cite{maldeniya2020samaritans}. 
Over time, the growth of rules can even evolve into bureaucracies~\cite{butler2008don,o2007emergence} or oligarchies~\cite{Shaw2014Oligarchy} and may need to evolve again to better retain new members~\cite{steinmacher2015social}.
Better monitoring capabilities could support communities in recognizing these challenges and allow them to pivot their governance to address them.

\subsection{Adversarial Usage}

Currently, PolicyKit requires cooperation between community members and admins on the platform because admins still have the ability to revoke PolicyKit's privileged access token, uninstalling it from the community. 
This vulnerability is a downside of software infrastructure for governance that exists separate from the technical permissions embedded in a platform and thus can never fully supersede it. 
However, this vulnerability exists whether communities use PolicyKit or not, as admins have the power to disregard rules or even delete the community.
In addition, platforms may have platform-wide policies such as a Terms of Service and a centralized process for enforcement. PolicyKit cannot and is not intended to override platform-wide policies of this nature. Instead, PolicyKit operates in the wide space of decisions above the relatively low bar that platforms set for unwanted behavior.

There are other forms of adversarial usage that may arise as communities use PolicyKit. One issue is loopholes, obscured by code or accidentally added, similar to text loopholes in complicated legal documents or the game Nomic~\cite{nomic}. Much like one needs a lawyer to parse legal documents, it may take technical expertise to discover loopholes in code.
How can communities recover from loopholes? In the worst case, if the entire government comes to a standstill, it should be possible to restart governance from the beginning or roll back a number of actions. Of course, these capabilities need to be designed so that they are not exploited as well.
A separate vulnerability arises from the PolicyKit system itself being hosted on a centralized webserver and database. Communities can use our hosted site or host their own instance, but in either case, access to the infrastructure is concentrated in a few hands. Novel distributed architectures or blockchain databases could potentially reduce this vulnerability in the future.

\section{Limitations and Future Work}

While we have demonstrated the expressivity of PolicyKit and its technical ability to govern communities, we have left to future work longitudinal field studies with online communities. As is, this work presents a low level framework upon which additional research and development is needed to open up policy-making capabilities for non-programmers and provide resources for communities to design good governance. We aim to work with a number of initial communities to understand what they wish to create as well as uncover usability issues.
These deployments with real communities will also allow us to co-design more sophisticated policies that can address real-world challenges that may arise, from the rise of politics or coordinated factions to loopholes and unrecoverable states.

\section{Conclusion}
In this work, we present PolicyKit, a novel software infrastructure that empowers online communities to succinctly author a broad range of governance models that can then be carried out on their home platform.
PolicyKit's expressiveness lies in its shift from the status quo of describing governance in terms of roles that are assigned permissions, towards articulating procedures for determining what is permissible.
We demonstrate through examples how procedures allow for the expression of a diversity of governance models, from elections, to reputation systems, to deliberative democracy.

In addition, the design of PolicyKit's infrastructure rests on the abstraction of governance as a series of actions that are successively proposed, and policies that are continually evaluating those actions. 
This abstraction allows us to simplify the execution of governance on users' home platforms so that writing a policy means only implementing several short functions.
It also enables not only everyday governance execution but also the gradual evolution of governance models by communities as they too evolve over time.

Taken altogether, PolicyKit significantly reduces barriers for communities to build their own governance. More powerfully, it instantiates a framework upon which additional systems can be built that continue to lower barriers to participation and broaden communities' governance capabilities.

\section{Acknowledgements}
Thank you to the Stanford HCI group and the MetaGov group for feedback, particularly Mark Whiting, Joseph Seering, Will Crichton, Evan Strasnick, Mitchell Gordon, and Seth Frey.

\balance{}

\bibliographystyle{SIGCHI-Reference-Format}
\bibliography{proceedings}


\begin{thebibliography}{00}


\ifx \showCODEN    \undefined \def \showCODEN     #1{\unskip}     \fi
\ifx \showDOI      \undefined \def \showDOI       #1{{\tt DOI:}\penalty0{#1}\ }
  \fi
\ifx \showISBNx    \undefined \def \showISBNx     #1{\unskip}     \fi
\ifx \showISBNxiii \undefined \def \showISBNxiii  #1{\unskip}     \fi
\ifx \showISSN     \undefined \def \showISSN      #1{\unskip}     \fi
\ifx \showLCCN     \undefined \def \showLCCN      #1{\unskip}     \fi
\ifx \shownote     \undefined \def \shownote      #1{#1}          \fi
\ifx \showarticletitle \undefined \def \showarticletitle #1{#1}   \fi
\ifx \showURL      \undefined \def \showURL       #1{#1}          \fi

\bibitem{alkhatib2019street}
{Ali Alkhatib} {and} {Michael Bernstein}. 2019.
\newblock \showarticletitle{Street-Level Algorithms: A Theory at the Gaps
  Between Policy and Decisions}. In {\em Proceedings of the 2019 CHI Conference
  on Human Factors in Computing Systems}. 1--13.
\newblock


\bibitem{scrapir}
{Tarfah Alrashed}, {Jumana Almahmoud}, {Amy~X. Zhang}, {and} {David Karger}.
  2020.
\newblock \showarticletitle{ScrAPIr: Making Web Data APIs Accessible to End
  Users}. In {\em Proceedings of the 2020 CHI conference on human factors in
  computing systems}.
\newblock


\bibitem{barlow1996Declaration}
{John~Perry Barlow}. 1996.
\newblock \showarticletitle{A Declaration of the Independence of Cyberspace}.
\newblock  (1996).
\newblock
\showURL{%
\url{https://projects.eff.org/~barlow/Declaration-Final.html}}


\bibitem{bryant2005becoming}
{Susan~L Bryant}, {Andrea Forte}, {and} {Amy Bruckman}. 2005.
\newblock \showarticletitle{Becoming Wikipedian: transformation of
  participation in a collaborative online encyclopedia}. In {\em Proceedings of
  the 2005 international ACM SIGGROUP conference on Supporting group work}.
  1--10.
\newblock


\bibitem{burke2008mopping}
{Moira Burke} {and} {Robert Kraut}. 2008.
\newblock \showarticletitle{Mopping up: modeling wikipedia promotion
  decisions}. In {\em Proceedings of the 2008 ACM conference on Computer
  supported cooperative work}. 27--36.
\newblock


\bibitem{butler2008don}
{Brian Butler}, {Elisabeth Joyce}, {and} {Jacqueline Pike}. 2008.
\newblock \showarticletitle{Don't look now, but we've created a bureaucracy:
  the nature and roles of policies and rules in wikipedia}. In {\em Proceedings
  of the SIGCHI conference on human factors in computing systems}. 1101--1110.
\newblock


\bibitem{centivany2016popcorn}
{Alissa Centivany} {and} {Bobby Glushko}. 2016.
\newblock \showarticletitle{"Popcorn Tastes Good" Participatory Policymaking
  and Reddit's "AMAgeddon"}. In {\em Proceedings of the 2016 CHI Conference on
  Human Factors in Computing Systems}. 1126--1137.
\newblock


\bibitem{chandrasekharan2019crossmod}
{Eshwar Chandrasekharan}, {Chaitrali Gandhi}, {Matthew~Wortley Mustelier},
  {and} {Eric Gilbert}. 2019.
\newblock \showarticletitle{Crossmod: A Cross-Community Learning-based System
  to Assist Reddit Moderators}.
\newblock {\em Proceedings of the ACM on Human-Computer Interaction\/} {3},
  CSCW (2019), 1--30.
\newblock


\bibitem{cohen2007contributor}
{Noam Cohen}. 2007.
\newblock \showarticletitle{A contributor to Wikipedia has his fictional side}.
\newblock {\em New York Times\/}  {5} (2007), 5.
\newblock


\bibitem{de2003democracy}
{Alexis De~Tocqueville}. 2003.
\newblock {\em Democracy in america}. Vol.~10.
\newblock Regnery Publishing.
\newblock


\bibitem{dibbell1994rape}
{Julian Dibbell}. 1994.
\newblock \showarticletitle{A rape in cyberspace or how an evil clown, a
  Haitian trickster spirit, two wizards, and a cast of dozens turned a database
  into a society}.
\newblock {\em Ann. Surv. Am. L.\/} (1994), 471.
\newblock


\bibitem{dosono2019moderation}
{Bryan Dosono} {and} {Bryan Semaan}. 2019.
\newblock \showarticletitle{Moderation practices as emotional labor in
  sustaining online communities: The case of AAPI identity work on Reddit}. In
  {\em Proceedings of the 2019 CHI Conference on Human Factors in Computing
  Systems}. 1--13.
\newblock


\bibitem{juries}
{Jenny Fan} {and} {Amy~X Zhang}. 2020.
\newblock \showarticletitle{Digital Juries: A Civics-Oriented Approach to
  Platform Governance}. In {\em Proceedings of the 2020 CHI Conference on Human
  Factors in Computing Systems}. 1--14.
\newblock


\bibitem{faridani2010opinion}
{Siamak Faridani}, {Ephrat Bitton}, {Kimiko Ryokai}, {and} {Ken Goldberg}.
  2010.
\newblock \showarticletitle{Opinion space: a scalable tool for browsing online
  comments}. In {\em Proceedings of the SIGCHI Conference on Human Factors in
  Computing Systems}. 1175--1184.
\newblock


\bibitem{farina2013democratic}
{Cynthia Farina}, {Hoi Kong}, {Cheryl Blake}, {Mary Newhart}, {and} {Nik Luka}.
  2013.
\newblock \showarticletitle{Democratic deliberation in the wild: The McGill
  online design studio and the RegulationRoom project}.
\newblock {\em Fordham Urb. LJ\/}  {41} (2013), 1527.
\newblock


\bibitem{fish2011birds}
{Adam Fish}, {Luis~FR Murillo}, {Lilly Nguyen}, {Aaron Panofsky}, {and}
  {Christopher~M Kelty}. 2011.
\newblock \showarticletitle{Birds of the Internet: Towards a field guide to the
  organization and governance of participation}.
\newblock {\em Journal of Cultural Economy\/} {4}, 2 (2011), 157--187.
\newblock


\bibitem{forte2009decentralization}
{Andrea Forte}, {Vanesa Larco}, {and} {Amy Bruckman}. 2009.
\newblock \showarticletitle{Decentralization in Wikipedia governance}.
\newblock {\em Journal of Management Information Systems\/} {26}, 1 (2009),
  49--72.
\newblock


\bibitem{nodejs}
{OpenJS Foundation}. 2020.
\newblock {\em Project Governance}.
\newblock
\showURL{%
\url{https://nodejs.org/en/about/governance/}}


\bibitem{freeman1972tyranny}
{Jo Freeman}. 1972.
\newblock \showarticletitle{The tyranny of structurelessness}.
\newblock {\em Berkeley Journal of Sociology\/} (1972), 151--164.
\newblock


\bibitem{frey2019place}
{Seth Frey}, {PM Krafft}, {and} {Brian~C Keegan}. 2019.
\newblock \showarticletitle{" This Place Does What It Was Built For" Designing
  Digital Institutions for Participatory Change}.
\newblock {\em Proceedings of the ACM on Human-Computer Interaction\/} {3},
  CSCW (2019), 1--31.
\newblock


\bibitem{frey2019emergence}
{Seth Frey} {and} {Robert~W Sumner}. 2019.
\newblock \showarticletitle{Emergence of integrated institutions in a large
  population of self-governing communities}.
\newblock {\em PloS one\/} {14}, 7 (2019), e0216335.
\newblock


\bibitem{geiger2014bots}
{R~Stuart Geiger}. 2014.
\newblock \showarticletitle{Bots, bespoke, code and the materiality of software
  platforms}.
\newblock {\em Information, Communication \& Society\/} {17}, 3 (2014),
  342--356.
\newblock


\bibitem{geiger2016bot}
{R~Stuart Geiger}. 2016.
\newblock \showarticletitle{Bot-based collective blocklists in Twitter: the
  counterpublic moderation of harassment in a networked public space}.
\newblock {\em Information, Communication \& Society\/} {19}, 6 (2016),
  787--803.
\newblock


\bibitem{geiger2010work}
{R~Stuart Geiger} {and} {David Ribes}. 2010.
\newblock \showarticletitle{The work of sustaining order in wikipedia: the
  banning of a vandal}. In {\em Proceedings of the 2010 ACM conference on
  Computer supported cooperative work}. 117--126.
\newblock


\bibitem{gillespie2018custodians}
{Tarleton Gillespie}. 2018.
\newblock {\em Custodians of the Internet: Platforms, content moderation, and
  the hidden decisions that shape social media}.
\newblock Yale University Press.
\newblock


\bibitem{grudin1994groupware}
{Jonathan Grudin}. 1994.
\newblock \showarticletitle{Groupware and social dynamics: Eight challenges for
  developers}.
\newblock {\it Commun. ACM} {37}, 1 (1994), 92--105.
\newblock


\bibitem{hardt2015google}
{Steve Hardt} {and} {Lia~CR Lopes}. 2015.
\newblock \showarticletitle{Google votes: A liquid democracy experiment on a
  corporate social network}.
\newblock  (2015).
\newblock


\bibitem{hirschman1970exit}
{Albert~O Hirschman}. 1970.
\newblock {\em Exit, voice, and loyalty: Responses to decline in firms,
  organizations, and states}. Vol.~25.
\newblock Harvard university press.
\newblock


\bibitem{im2018deliberation}
{Jane Im}, {Amy~X Zhang}, {Christopher~J Schilling}, {and} {David Karger}.
  2018.
\newblock \showarticletitle{Deliberation and Resolution on Wikipedia: A Case
  Study of Requests for Comments}.
\newblock {\em Proceedings of the ACM on Human-Computer Interaction\/} {2},
  CSCW (2018), 1--24.
\newblock


\bibitem{jhaver2019human}
{Shagun Jhaver}, {Iris Birman}, {Eric Gilbert}, {and} {Amy Bruckman}. 2019a.
\newblock \showarticletitle{Human-machine collaboration for content regulation:
  The case of Reddit Automoderator}.
\newblock {\em ACM Transactions on Computer-Human Interaction (TOCHI)\/} {26},
  5 (2019), 1--35.
\newblock


\bibitem{jhaver2019does}
{Shagun Jhaver}, {Amy Bruckman}, {and} {Eric Gilbert}. 2019b.
\newblock \showarticletitle{Does transparency in moderation really matter? User
  behavior after content removal explanations on reddit}.
\newblock {\em Proceedings of the ACM on Human-Computer Interaction\/} {3},
  CSCW (2019), 1--27.
\newblock


\bibitem{jhaver2018online}
{Shagun Jhaver}, {Sucheta Ghoshal}, {Amy Bruckman}, {and} {Eric Gilbert}. 2018.
\newblock \showarticletitle{Online harassment and content moderation: The case
  of blocklists}.
\newblock {\em ACM Transactions on Computer-Human Interaction (TOCHI)\/} {25},
  2 (2018), 1--33.
\newblock


\bibitem{perspective}
{Google Jigsaw}. 2017.
\newblock Perspective API.
\newblock   (2017).
\newblock
\showURL{%
Retrieved May 5, 2020 from \url{https://www.perspectiveapi.com}}


\bibitem{kelty2018two}
{Christopher Kelty} {and} {Seth Erickson}. 2018.
\newblock \showarticletitle{Two modes of participation: A conceptual analysis
  of 102 cases of Internet and social media participation from 2005--2015}.
\newblock {\em The Information Society\/} {34}, 2 (2018), 71--87.
\newblock


\bibitem{kiene2016surviving}
{Charles Kiene}, {Andr{\'e}s Monroy-Hern{\'a}ndez}, {and} {Benjamin~Mako Hill}.
  2016.
\newblock \showarticletitle{Surviving an eternal september: How an online
  community managed a surge of newcomers}. In {\em CHI}. ACM, 1152--1156.
\newblock


\bibitem{kiser1982three}
{Larry Kiser} {and} {Elinor Ostrom}. 1982.
\newblock \showarticletitle{The three worlds of political action}.
\newblock {\em Strategies of political inquiry. Berverly Hilly: Sage\/} (1982).
\newblock


\bibitem{klein2011harvest}
{Mark Klein}. 2011.
\newblock \showarticletitle{How to harvest collective wisdom on complex
  problems: An introduction to the mit deliberatorium}.
\newblock {\em Center for Collective Intelligence working paper\/} (2011).
\newblock


\bibitem{konieczny2009governance}
{Piotr Konieczny}. 2009.
\newblock \showarticletitle{Governance, organization, and democracy on the
  Internet: The iron law and the evolution of Wikipedia}. In {\em Sociological
  Forum}, Vol.~24. Wiley Online Library, 162--192.
\newblock


\bibitem{konieczny2018volunteer}
{Piotr Konieczny}. 2018.
\newblock \showarticletitle{Volunteer retention, burnout and dropout in online
  voluntary organizations: stress, conflict and retirement of Wikipedians}.
\newblock {\em Research in Social Movements, Conflicts and Change (Volume 42).
  Emerald Publishing Limited\/} (2018), 199--219.
\newblock


\bibitem{kou2017managing}
{Yubo Kou}, {Xinning Gui}, {Shaozeng Zhang}, {and} {Bonnie Nardi}. 2017.
\newblock \showarticletitle{Managing disruptive behavior through
  non-hierarchical governance: Crowdsourcing in League of Legends and Weibo}.
\newblock {\em Proceedings of the ACM on Human-Computer Interaction\/} {1},
  CSCW (2017), 1--17.
\newblock


\bibitem{kou2014governance}
{Yubo Kou} {and} {Bonnie~A Nardi}. 2014.
\newblock \showarticletitle{Governance in League of Legends: A hybrid system}.
  In {\em Foundations of Digital Games}.
\newblock


\bibitem{kraut2012building}
{Robert~E Kraut} {and} {Paul Resnick}. 2012.
\newblock {\em Building successful online communities: Evidence-based social
  design}.
\newblock Mit Press.
\newblock


\bibitem{kriplean2012supporting}
{Travis Kriplean}, {Jonathan Morgan}, {Deen Freelon}, {Alan Borning}, {and}
  {Lance Bennett}. 2012a.
\newblock \showarticletitle{Supporting reflective public thought with
  considerit}. In {\em Proceedings of the ACM 2012 conference on Computer
  Supported Cooperative Work}. 265--274.
\newblock


\bibitem{kriplean2012you}
{Travis Kriplean}, {Michael Toomim}, {Jonathan Morgan}, {Alan Borning}, {and}
  {Andrew Ko}. 2012b.
\newblock \showarticletitle{Is this what you meant? Promoting listening on the
  web with reflect}. In {\em Proceedings of the SIGCHI Conference on Human
  Factors in Computing Systems}. 1559--1568.
\newblock


\bibitem{lampe2004slash}
{Cliff Lampe} {and} {Paul Resnick}. 2004.
\newblock \showarticletitle{Slash (dot) and burn: distributed moderation in a
  large online conversation space}. In {\em Proceedings of the SIGCHI
  conference on Human factors in computing systems}. 543--550.
\newblock


\bibitem{mahar2018squadbox}
{Kaitlin Mahar}, {Amy~X Zhang}, {and} {David Karger}. 2018.
\newblock \showarticletitle{Squadbox: A tool to combat email harassment using
  friendsourced moderation}. In {\em Proceedings of the 2018 CHI Conference on
  Human Factors in Computing Systems}. 1--13.
\newblock


\bibitem{maldeniya2020samaritans}
{Danaja Maldeniya}, {Ceren Budak}, {Lionel Robert}, {and} {Daniel Romero}.
  2020.
\newblock \showarticletitle{Herding a Deluge of Good Samaritans: How GitHub
  Projects Respond to Increased Attention}.
\newblock
\showDOI{%
\url{http://dx.doi.org/10.1145/3366423.3380272}}


\bibitem{malone2007harnessing}
{Thomas~W Malone} {and} {Mark Klein}. 2007.
\newblock \showarticletitle{Harnessing collective intelligence to address
  global climate change}.
\newblock {\em Innovations: Technology, Governance, Globalization\/} {2}, 3
  (2007), 15--26.
\newblock


\bibitem{mamykina2011design}
{Lena Mamykina}, {Bella Manoim}, {Manas Mittal}, {George Hripcsak}, {and}
  {Bj{\"o}rn Hartmann}. 2011.
\newblock \showarticletitle{Design lessons from the fastest q\&a site in the
  west}. In {\em Proceedings of the SIGCHI conference on Human factors in
  computing systems}. 2857--2866.
\newblock


\bibitem{matias2016going}
{J~Nathan Matias}. 2016.
\newblock \showarticletitle{Going dark: Social factors in collective action
  against platform operators in the Reddit blackout}. In {\em Proceedings of
  the 2016 CHI conference on human factors in computing systems}. 1138--1151.
\newblock


\bibitem{matias2019preventing}
{J~Nathan Matias}. 2019.
\newblock \showarticletitle{Preventing harassment and increasing group
  participation through social norms in 2,190 online science discussions}.
\newblock {\em Proceedings of the National Academy of Sciences\/} {116}, 20
  (2019), 9785--9789.
\newblock


\bibitem{matias2018civilservant}
{J~Nathan Matias} {and} {Merry Mou}. 2018.
\newblock \showarticletitle{CivilServant: Community-led experiments in platform
  governance}. In {\em Proceedings of the 2018 CHI conference on human factors
  in computing systems}. 1--13.
\newblock


\bibitem{mcginnis2011introduction}
{Michael~D McGinnis}. 2011.
\newblock \showarticletitle{An introduction to IAD and the language of the
  Ostrom workshop: a simple guide to a complex framework}.
\newblock {\em Policy Studies Journal\/} {39}, 1 (2011), 169--183.
\newblock


\bibitem{mcginnis2014social}
{Michael~D McGinnis} {and} {Elinor Ostrom}. 2014.
\newblock \showarticletitle{Social-ecological system framework: initial changes
  and continuing challenges}.
\newblock {\em Ecology and Society\/} {19}, 2 (2014).
\newblock


\bibitem{mnookin1996virtual}
{Jennifer~L Mnookin}. 1996.
\newblock \showarticletitle{Virtual (ly) law: The emergence of law in
  LambdaMoo: Mnookin}.
\newblock {\em Journal of computer-mediated communication\/} {2}, 1 (1996),
  JCMC214.
\newblock


\bibitem{muller2013work}
{Claudia M{\"u}ller-Birn}, {Leonhard Dobusch}, {and} {James~D Herbsleb}. 2013.
\newblock \showarticletitle{Work-to-rule: the emergence of algorithmic
  governance in Wikipedia}. In {\em Proceedings of the 6th International
  Conference on Communities and Technologies}. 80--89.
\newblock


\bibitem{myers2000past}
{Brad Myers}, {Scott~E Hudson}, {and} {Randy Pausch}. 2000.
\newblock \showarticletitle{Past, present, and future of user interface
  software tools}.
\newblock {\em ACM Transactions on Computer-Human Interaction (TOCHI)\/} {7}, 1
  (2000), 3--28.
\newblock


\bibitem{noveck2009wiki}
{Beth~Simone Noveck}. 2009.
\newblock {\em Wiki government: How technology can make government better,
  democracy stronger, and citizens more powerful}.
\newblock Brookings Institution Press.
\newblock


\bibitem{o2007emergence}
{Siobh{\'a}n O'mahony} {and} {Fabrizio Ferraro}. 2007.
\newblock \showarticletitle{The emergence of governance in an open source
  community}.
\newblock {\em Academy of Management Journal\/} {50}, 5 (2007), 1079--1106.
\newblock


\bibitem{ostrom}
{Elinor Ostrom}. 1990.
\newblock {\em Governing the commons-The evolution of institutions for
  collective actions}.
\newblock Political economy of institutions and decisions.
\newblock


\bibitem{ransbotham2011membership}
{Sam Ransbotham} {and} {Gerald~C Kane}. 2011.
\newblock \showarticletitle{Membership turnover and collaboration success in
  online communities: Explaining rises and falls from grace in Wikipedia}.
\newblock {\em Mis Quarterly\/} (2011), 613--627.
\newblock


\bibitem{raymond1998homesteading}
{Eric~S Raymond}. 1998.
\newblock \showarticletitle{Homesteading the noosphere}.
\newblock  (1998).
\newblock


\bibitem{roberts2019behind}
{Sarah~T Roberts}. 2019.
\newblock {\em Behind the screen: Content moderation in the shadows of social
  media}.
\newblock Yale University Press.
\newblock


\bibitem{salehi2015we}
{Niloufar Salehi}, {Lilly~C Irani}, {Michael~S Bernstein}, {Ali Alkhatib}, {Eva
  Ogbe}, {and} {Kristy Milland}. 2015.
\newblock \showarticletitle{We are dynamo: Overcoming stalling and friction in
  collective action for crowd workers}. In {\em Proceedings of the 33rd annual
  ACM conference on human factors in computing systems}. 1621--1630.
\newblock


\bibitem{schneider-admins2019}
{Nathan Schneider}. 2019.
\newblock Admins, {{Mods}}, and {{Benevolent Dictators}} for {{Life}}: {{The
  Implicit Feudalism}} of {{Online Communities}}.  (2019).
\newblock
\showURL{%
\url{https://ntnsndr.in/implicit-feudalism}}


\bibitem{schoenebeck2020drawing}
{Sarita Schoenebeck}, {Oliver~L Haimson}, {and} {Lisa Nakamura}. 2020.
\newblock \showarticletitle{Drawing from justice theories to support targets of
  online harassment}.
\newblock {\em new media \& society\/} (2020), 1461444820913122.
\newblock


\bibitem{Shaw2014Oligarchy}
{Aaron Shaw} {and} {Benjamin~M. Hill}. 2014.
\newblock \showarticletitle{{Laboratories of Oligarchy? How the Iron Law
  Extends to Peer Production}}.
\newblock {\em Journal of Communication\/} {64}, 2 (03 2014), 215--238.
\newblock


\bibitem{steinmacher2015social}
{Igor Steinmacher}, {Tayana Conte}, {Marco~Aur{\'e}lio Gerosa}, {and} {David
  Redmiles}. 2015.
\newblock \showarticletitle{Social barriers faced by newcomers placing their
  first contribution in open source software projects}. In {\em Proceedings of
  the 18th ACM conference on Computer supported cooperative work \& social
  computing}. 1379--1392.
\newblock


\bibitem{nomic}
{Peter Suber}. 1990.
\newblock \showarticletitle{Appendix 3: {{Nomic}}: {{A Game}} of
  {{Self}}-{{Amendment}}}.
\newblock In {\em The {{Paradox}} of {{Self}}-{{Amendment}}: {{A Study}} of
  {{Law}}, {{Logic}}, {{Omnipotence}}, and {{Change}}}. {Peter Lang}.
\newblock
\showISBNx{978-0-8204-1212-2}
\showURL{%
\url{https://dash.harvard.edu/handle/1/10288408}}


\bibitem{zhang2017wikum}
{Amy~X Zhang}, {Lea Verou}, {and} {David Karger}. 2017.
\newblock \showarticletitle{Wikum: Bridging discussion forums and wikis using
  recursive summarization}. In {\em Proceedings of the 2017 ACM Conference on
  Computer Supported Cooperative Work and Social Computing}. 2082--2096.
\newblock


\bibitem{zheng2019roles}
{Lei Zheng}, {Christopher~M Albano}, {Neev~M Vora}, {Feng Mai}, {and}
  {Jeffrey~V Nickerson}. 2019.
\newblock \showarticletitle{The Roles Bots Play in Wikipedia}.
\newblock {\em Proceedings of the ACM on Human-Computer Interaction\/} {3},
  CSCW (2019), 1--20.
\newblock


\end{thebibliography}

\end{document}